\documentclass[aps,pra,showpacs,preprintnumbers,amsmath,amssymb,footinbib,notitlepage]{revtex4-1}

\usepackage[utf8]{inputenc}
\usepackage{hyperref}
\usepackage{listings}
\usepackage{xcolor}
\usepackage{amssymb}
\usepackage{amsmath}
\usepackage{graphicx}
\usepackage{subcaption}
\usepackage{url}
\graphicspath{ {./images/} }

\definecolor{codegreen}{rgb}{0,0.6,0}
\definecolor{codegray}{rgb}{0.5,0.5,0.5}
\definecolor{codepurple}{rgb}{0.58,0,0.82}
\definecolor{backcolour}{rgb}{0.95,0.95,0.92}

\lstdefinestyle{mystyle}{
    backgroundcolor=\color{backcolour},
    commentstyle=\color{codegreen},
    keywordstyle=\color{magenta},
    numberstyle=\tiny\color{codegray},
    stringstyle=\color{codepurple},
    basicstyle=\ttfamily\footnotesize,
    breakatwhitespace=false,
    breaklines=true,
    captionpos=b,
    keepspaces=true,
    numbers=left,
    numbersep=5pt,
    showspaces=false,                
    showstringspaces=false,
    showtabs=false,
    tabsize=2
}

\begin{document}




\title{ML4Chem: A Machine Learning Package for Chemistry and Materials Science.}

\author{Muammar El Khatib}
 \email{melkhatibr@lbl.gov}
\affiliation{%
Computational Research Division, Lawrence Berkeley National Laboratory, Berkeley, CA 94720, USA
}%
\author{Wibe A de Jong}%
\affiliation{%
Computational Research Division, Lawrence Berkeley National Laboratory, Berkeley, CA 94720, USA
}%

\date{\today}

\begin{abstract} 

\textsf{ML4Chem} is an open-source machine learning library for chemistry and
materials science. It provides an extendable platform to develop and deploy machine
learning models and pipelines and is targeted to the non-expert and expert users.
\textsf{ML4Chem} follows user-experience design and offers the needed tools
to go from data preparation to inference. Here we introduce
its \lstinline{atomistic} module for the implementation, deployment, and
reproducibility of atom-centered models. This module is composed of six core
building blocks: data, featurization, models, model optimization,
inference, and visualization. We present their functionality and easiness 
of use with demonstrations utilizing neural networks and kernel
ridge regression algorithms.
\end{abstract}

\maketitle

\section{Introduction}

In the last decade, machine learning (ML) has undergone fast development due
to large amounts of available data and advancements in computational hardware
\textit{e.g.} faster and cheaper central processing units (CPU), graphics
process units (GPU), and more recently the introduction of tensor processing
units (TPU). Algorithmic improvements on how to compute the gradient in
weight space of feedforward neural networks with respect to a loss
function\cite{Goodfellow2016} reduced the computational time of training deep
neural networks significantly. As a result companies like Google, and
Facebook, introduced the most useful deep learning platforms available right
now: TensorFlow\cite{TensorFlow}, and Pytorch\cite{NEURIPS2019_9015}. These
frameworks positively impacted and advanced ML research because they helped
with democratizing and simplified access to ML technologies to a larger
audience.

In the field of physical chemistry and materials sciences, ML models are
being standardized and applied to solve tasks such as the acceleration of
atomistic
simulations\cite{Artrith2012,Artrith2011,Behler2010,Behler2007,Pilania2013},
prediction of the electronic Hamiltonian with generative
models\cite{Toth2019,Schutt2019a}, extraction of continuous latent
representations for the generation of molecules\cite{Gomez-Bombarelli2018},
and even the prediction of the scent of small organic
molecules\cite{Sanchez-Lengeling2019}. It also is becoming the norm to
release software solutions as support to validate results of publications
that apply ML models, and alleviate the ``reproducibility crisis in
artificial intelligence and machine learning''\cite{Hutson2018, Stupple2019}.
Nevertheless, this obliquely fragments the software ecosystem because each
software implementation \textit{a)} requires specific data structures and
\textit{b)} would likely experience a lack of continuous support. 
There already are packages that democratize ML in chemistry. For example,
DeepChem\cite{deepchem} has played a critical role in providing users 
a helpful platform of ML algorithms and featurizers for drug discovery, quantum
chemistry, material sciences, and biology. More recently ChemML has been
introduced as a machine learning and informatics program suite for the
analysis, mining, and modeling of chemical and materials
data\cite{chemml2019}. What differentiates \textsf{ML4Chem} is that it 
focuses on easing the implementation of new functionality, extraction of 
intermediate quantities, interfacing with external programs, and exportation 
of any of its modules' outputs. Also, \textsf{ML4Chem} is in its infancy bringing 
up the possibility to shape its future directions based on current users' 
needs and ML paradigms.

Here we introduce the \lstinline{atomistic} module where
ML algorithms learn underlying relationships between molecules and properties
treating atoms as central objects. They exploit the principle of locality in
Physics: \textit{a global quantity is defined as a sum that runs over many
localized contributions}. These localized contributions usually account for
interactions of an atom and its nearest-neighbor atoms (many-body
interactions). Atomistic models are very useful and have been successfully 
applied for the acceleration of molecular dynamics
simulations\cite{Li2015,Botu2015b,Chmiela2017}, identification of phase
transitions in materials\cite{Li2018a}, determination of energy and atomic
forces with high accuracy\cite{Botu2015,Botu2017a}, the search of
saddle-points\cite{Peterson2016a} and the prediction of atomic
charges\cite{Ghasemi2015,Faraji2017}.

This publication is organized as follows: in
section~\ref{sec:atomisticdesignandarch}, we will discuss the design and
architecture of \textsf{ML4Chem}'s \lstinline{atomistic} module. Each of its
core blocks is introduced in Section~\ref{sec:coremodules} and we will
demonstrate the code's capabilities through a series of demonstration
examples in Section~\ref{sec:demonstrations}. Finally, conclusions and
perspectives are drawn.

\section{Atomistic Module: Design and Architecture}
\label{sec:atomisticdesignandarch}

\textsf{ML4Chem} and its modules are written in Python in an object-oriented
programming paradigm and are built on top of popular open-source projects to
avoid duplication of efforts. In this regard, all deep learning computations
are implemented with Pytorch\cite{NEURIPS2019_9015}. Mathematical and linear
algebra operations are executed by Numpy\cite{VanderWalt2011} or
Scipy\cite{Millman2011, 2019arXiv190710121V} that are widely used and
recognized for this purpose. Parallelism is achieved with a flexible library
for parallel computing called Dask\cite{DaskDevelopmentTeam2016}. Dask
enables computational scaling-up from a laptop to High-Performance Computing
(HPC) clusters effortlessly and offers a web dashboard to real-time
monitoring. This is particularly valuable because it provides a good
estimation to users about the status of calculations, and helps at profiling
computations. Good documentation is another important aspect, as the lack of it can harm usability.
\textsf{ML4Chem}'s source code is documented using Numpy Python docstrings
and rendered in HTML and PDF format. Also, we provide diverse information ranging
from installation, theory, usage of modules, and examples.

\textsf{ML4Chem}'s modules are developed following user-experience (UX)
design practices to deliver usability, accessibility, and desirability. For
example, in \textsf{ML4Chem} the names of modules, classes, and functions
tend to be idiomatic and easy to remember semantically. Getting the latent
space from an autoencoder is performed by calling
\lstinline{autoencoder.get_latent_space(X)}, or the computation of atomistic
features is done with a \lstinline{.calculate()} class method that is
provided in all featurizers under the \lstinline{atomistic.features} module
\textit{e.g.} \lstinline{features.calculate(X, purpose="inference")}. All
modules are designed to have the same structure, enabling users to become
familiar and gain intuition on their usage quickly. The library can be used
in interactive Python environments such as iPython, Jupyter or JupyterLab
notebooks. Or if desired, as scripts that are invoked by the Python
interpreter.

\begin{figure}[ht]
    \centering
    \includegraphics[width=12.5cm]{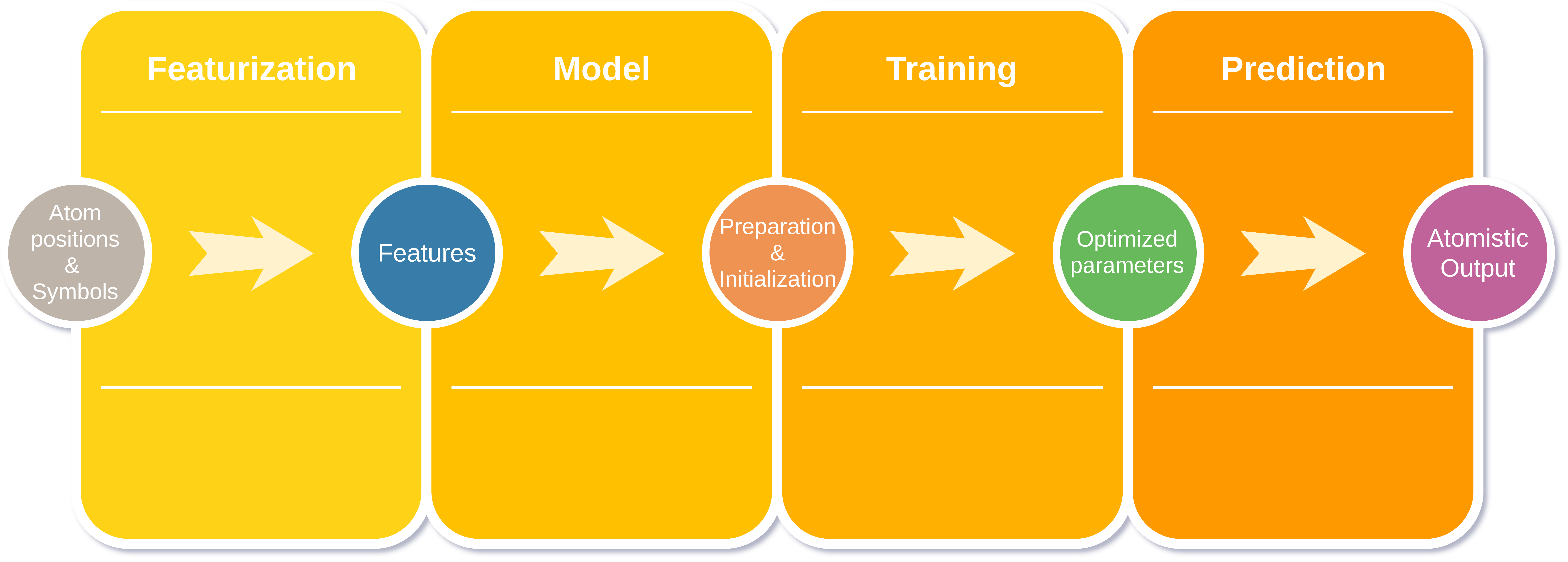}
    \caption{Design approach of the \lstinline{atomistic} module of \textsf{ML4Chem}.}
    \label{fig:atomistic_schema}
\end{figure}

Figure~\ref{fig:atomistic_schema} shows a schematic representation of the
machine learning workflow that drives the design philosophy used to develop
the \lstinline{atomistic} module in
\textsf{ML4Chem}:

\begin{enumerate}
    \item Atoms positions are mapped into atomistic feature vectors with the
    \lstinline{atomistic.features} module and chemical symbols are used as labels.
    \item ML models are instantiated utilizing atomistic feature vectors
    as input, and depending on the nature of the task (supervised or
    unsupervised learning), targets might or might not be known.
    \item Model's parameters are trained either by minimizing/maximizing a
    loss function or solving systems of equations. In the former case,
    \textsf{ML4Chem} provides \lstinline{loss} and \lstinline{optim} modules
    with sets of predefined loss functions and optimizers to train supervised
    and unsupervised atomistic models.
    \item The resulting atomistic model outputs and predictions are expected
    to be scalar or vector quantities.
\end{enumerate}

\begin{figure}[ht]
    \centering
    \includegraphics[width=10cm]{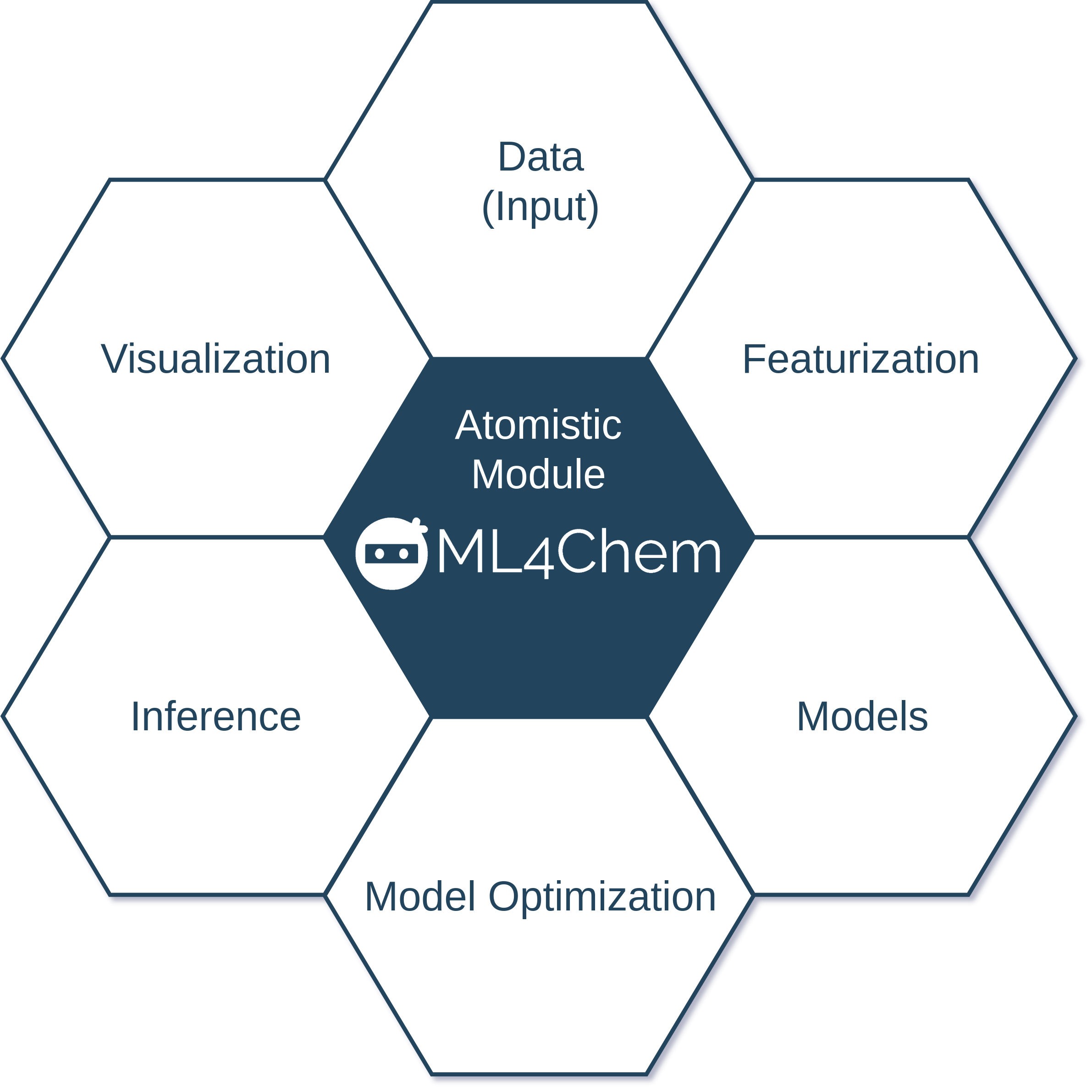}
    \caption{Core modules of the \lstinline{atomistic} module of \textsf{ML4Chem}.}
    \label{fig:modular}
\end{figure}

As shown in Figure~\ref{fig:modular}, the architecture of the
\lstinline{atomistic} module is composed of 6 building blocks. They
correspond to the methods and tools required to deploy atom-centered
simulations from input featurization to inference and visualization,
according to our design philosophy. Modules inside the code blocks
need to comply with being \textit{derived classes} from \textit{base classes}
using Python mixins. This guarantees new classes are reusing the code base
and inheriting already defined structures implicitly from the \textit{base
classes} to operate seamlessly with other \lstinline{ML4Chem} components.
This practice is encouraged by providing a \lstinline{base.py} file within
each module level that contains required base classes, and is enforced in the
continuous integration (CI) system.

In the following sections, each of these blocks is discussed with particular
attention on what can be achieved with them, their implementation, and code
snippets on their usage. When relevant, we will discuss the theory and
mathematics behind them as well.

\section{Core Modules}
\label{sec:coremodules}
\subsection{Data}
\label{sec:datahandling}

ML is focused on finding underlying patterns and relationships based on data
examples. The format of data, a central part of ML, varies depending on the
sources, and the ML algorithm adopted for solving a particular task.
\textsf{ML4Chem} provides a \lstinline{Data} class that creates an object
with a data structure that facilitates interoperability with any of its
available \lstinline{atomistic} modules.

Our \lstinline{atomistic} ML algorithms require molecules in the form of
\lstinline{Atoms} objects as implemented in the Atomic Simulation Environment
(ASE)\cite{HjorthLarsen2017}. One of the reasons behind choosing ASE is its
stability and that it supports more than 50 file formats in its
\lstinline{ase.io.read} module \textit{e.g.} XYZ, NWChem, GPAW, and Gaussian.
For instance, an XYZ file can be parsed and converted to an \lstinline{Atoms}
object in the following way: \lstinline{molecule = ase.io.read("file.xyz")}.
\lstinline{Atoms} objects hold molecular information such as Cartesian
coordinates, atom types, chemical symbols, molecular charge, cell type,
energy, and atomic forces. Molecules in ASE format can be stored to disk as a
list of molecules (\lstinline{Atoms}) using ASE's \lstinline{Trajectory}
module. Besides the formats supported by the \lstinline{ase.io.read} module,
we also can parse the Chemical JSON (CJSON) format\cite{Hanwell2017} and the
ANI-1 data set\cite{Smith2017} using the \lstinline{data.parser} module.
Support of input formats such as those available in pymatgen\cite{Ong2013}
and MolSSI's QCSchema are planned for future releases.

The \lstinline{Data} class uses a list of molecules, or in other words
\textit{a list \lstinline{Atoms} objects}, to generate a unique \textit{sha1}
hash to label each molecule and store their respective pairs input/targets.
Targets refer to the expected output of ML models and in
\lstinline{atomistic} simulations they may correspond to total energy, atomic
forces, dipole moments, etc. Duplicated data points are automatically removed
during the hashing procedure to avoid poor performance and numerical
instability. In Listing~\ref{lst:data}, the \lstinline{Data} class is
instantiated by passing an ASE trajectory file with name
\lstinline{dataset.traj} containing some \lstinline{molecules} for the
\lstinline{purpose} of \lstinline{"training"} an atomistic ML algorithm.
After hashing the \lstinline{molecules} present in the trajectory file, pairs
of input/targets examples are yielded by invoking the \lstinline{.get_data()}
class method and assigned to the \lstinline{training_set} and
\lstinline{targets} variables.

Finally, once the data is loaded in memory and arranged by the
\lstinline{Data} class, it can be mapped into features by the
\lstinline{atomistic.features} module of \textsf{ML4Chem}. The 
content of this object,
an ordered dictionary, can be easily exported as a pandas dataframe using the
\lstinline{.to_pandas()} class method and saved or serialized for subsequent
use.

\clearpage
\begin{lstlisting}[caption={Example of Data class usage in \textsf{ML4Chem}.},
                   label={lst:data},language=Python]
from ase.io import Trajectory
from ml4chem.data.handler import Data

molecules = Trajectory("dataset.traj")
purpose = "training"

data_handler = Data(molecules, purpose=purpose)
training_set, targets = data_handler.get_data(purpose=purpose)
df = data_handler.to_pandas()
\end{lstlisting}

\subsection{Featurization}

ML features are defined as a set of measurable unique characteristics or
properties of an observable. They are fundamental to any ML algorithm because
they represent what models ``see''. Feature engineering is the process of
applying domain knowledge to generate sets of numerical features that make ML
algorithms work and learn meaningful representations from data. In physics
constrained domains, like atomistic ML models, feature extraction is also
elusive because features require to fulfill a series of properties that are
commonly expected from physical systems such as rotational and translational
invariance (equivariance). Featurization is a challenging time-consuming
process and the feature engineering cycle encompasses their creation,
selection according to importance, and validation for the task of interest.

Features can be classified in \textit{i)} human-engineered features where
assumptions about data are made by humans to assign properties to observables
or \textit{ii)} machine-engineered features where the ML algorithms discover
meaningful representation during the training procedure. The
\lstinline{atomistic.features} module of \textsf{ML4Chem} supports the former
case with Gaussian type features, and the latter with a
\lstinline{LatentFeatures} class (introduced in the following subsections).

To avoid duplication of efforts, atomistic features such as Coulomb
matrix\cite{Rupp2012}, smooth overlap of atomic positions
(SOAP)\cite{Bartok2013}, and many-body tensor representation
(MBTR)\cite{Huo2017a} are supported through DScribe which is a software
package for ML that provides popular feature transformations
(``descriptors'') for atomistic materials simulations. This accelerates the
application of ML for atomistic property prediction by
providing a user-friendly, off-the-shelf descriptor
implementations\cite{Himanen2019}. Our own implementation of the Coulomb matrix
feature vectors is available in the \lstinline{atomistic.features} module 
of \textsf{ML4Chem}, and serves as an example of how easily our package 
can be extended.

\subsubsection{Gaussian Features}

In 2007, Behler and Parrinello\cite{Behler2007} introduced Gaussian feature
vectors, also referred to as ``symmetry functions'' (SF), for the
representation of high-dimensional potential energy surfaces with artificial
neural networks. These features overcome limitations related to the
\textit{image-centered} models and are built for each atom in a molecule or
extended system. They \textit{fingerprint} the \textbf{relevant chemical
environment} of atoms in molecules, and their computation only requires
chemical symbols and atomic positions.

To delimit the {\em effective range} of interactions within the domain of a
central atom, a {\em cutoff function} ($f_c$) is introduced:

\begin{equation} \label{eq:cutofff}
    f_c(r) =
    \begin{cases}
        0.5(1+cos(\pi \frac{r}{R_c}))   , & \text{if}\ r \leq R_c, \\
                                    0   , & \text{if}\ r \geq R_c.
    \end{cases}
\end{equation}

Where $R_c$ is the cutoff radius (in unit length), and $r$ is the
inter-atomic distance between atoms $i$ and $j$. The cutoff function, with
Cosine shape as shown in Eq.~\ref{eq:cutofff}, vanishes for inter-atomic
separations larger than $R_c$ and takes finite values below the cutoff radius.
These cutoff functions aim to avoid abrupt changes in the magnitudes of the
features near the boundary by smoothly damping them.

There are two sets of interactions to consider when building Gaussian
features: \textit{i)} the radial (two-body term) and \textit{ii)} angular
(three-body terms) SFs. The radial SFs account for all possible interactions
between a central atom $i$ and its nearest neighbors atoms $j$. It is defined
by Eq.~\ref{eq:g1sf}:

\begin{equation} \label{eq:g1sf}
    \mathbf{G_i^2} = \sum_{j = 1}^{N_{atom}} e^{-\eta(\mathbf{R_{ij}}-R_{s})^2/R_c^2} f_c(R_{ij}),
\end{equation}

where $\mathbf{R_{ij}}$ is the Euclidean distance between central atom $i$
and neighbor atom $j$, $R_s$ defines the center of the Gaussian, and $\eta$
is related to its width. Each pairwise contribution to the feature in the sum
is normalized by the square of the cutoff radius
$R_c^2$ as proposed in Ref.~\cite{Khorshidi2016}. In practice, one builds a high-dimensional
feature vector by choosing different $\eta$ values.

In addition to the radial SFs (two-body term), it is possible to include
triplet many-body interactions within the cutoff radius $R_c$ using the
following equation:

\begin{equation} \label{eq:g2sf}
    \mathbf{G_i^3} = 2^{1-\zeta} \sum_{j, k \neq i} (1 + \lambda cos\theta_{ijk})^{\zeta} e^{-\eta
        (\mathbf{R_{ij}}^2 + \mathbf{R_{ik}}^2 + \mathbf{R_{jk}}^2)/R_c^2} f_c(R_{ij}) f_c(R_{ik})
        f_c(R_{jk}).
\end{equation}

This part of the feature vector is built from considering the Cosine between all
possible $\theta_{ijk}$ angles of a central atom $i$ and a pair of neighbors
$j$, and $k$. There exists a variant of $\mathbf{G_i^3}$ that includes
three-body interactions of triplets forming $180^{\circ}$ inside the cutoff
sphere but having an inter-atomic separation larger than $Rc$. These SFs
account for long-range interactions as described by Behler in
Ref.~\cite{Behler2015}:

\begin{equation}\label{eq:g3sf}
    \mathbf{G_i^4} = 2^{1-\zeta} \sum_{j, k \neq i} (1 + \lambda cos\theta_{ijk})^{\zeta} e^{-\eta
        (\mathbf{R_{ij}}^2 + \mathbf{R_{ik}}^2)/R_c^2} f_c(R_{ij}) f_c(R_{ik}).
\end{equation}

In \textsf{ML4Chem}, Gaussian features can be built with the
\lstinline{atomistic.features} module as shown in Listing~\ref{lst:gaussian}.
The \lstinline{Gaussian()} class is instantiated with the desired cutoff
radius (units are \AA) to define the neighbor atoms, the type of angular
symmetry functions (either $\mathbf{G_i^3}$ or $\mathbf{G_i^4}$), and we
normalized dividing them by $R_c^2$. It also is possible to pass the
\lstinline{svm} keyword argument to calculate features for SVM algorithms. Note
that we need to pass the \lstinline{data_handler} and
\lstinline{training_set} objects created by the \lstinline{Data} class (see
Listing~\ref{lst:data}). It is important to preprocess, scale and normalize
features. In this way, models learn meaningful and noiseless underlying
representations. \textsf{ML4Chem} uses scikit-learn\cite{scikit-learn} for
the preprocessing of atomistic features. This can be activated by passing the
\lstinline{preprocessor} keyword argument. Currently, the supported
preprocessors in \textsf{ML4Chem} for atomistic features are:
\lstinline{MinMaxScaler}, \lstinline{StandardScaler}, and
\lstinline{Normalizer}. Finally, the \lstinline{.calculate()} method
calculates features.

\begin{lstlisting}[caption={Computing Gaussian
features with the \lstinline{atomistic.features} module of
\textsf{ML4Chem}.}, label={lst:gaussian},language=Python]
from ml4chem.atomistic.features import Gaussian

features = Gaussian(
    cutoff=6.5,
    normalized=True,
    preprocessor="MaxMinScaler",
    save_preprocessor="features.scaler",
    angular_type="G3",
)

X = features.calculate(
    training_set, purpose="training", data=data_handler, svm=False
)
\end{lstlisting}

\subsubsection{Latent Features}

In deep learning, latent features are non-directly observed variables
inferred by causality\cite{Borsboom2003}. These features fall under the
machine-engineered classification and are determined by the ML algorithm itself
during training \textit{without human intervention}. A clear example would
correspond to the informational bottleneck inferred when training
autoencoders (AE, see Section~\ref{sec:unsupervisedlearning}). This
informational bottleneck encodes hidden information by making a
dimensionality reduction that can reconstruct the input space (directly
observed variables). In physics constrained ML models, latent spaces might
correspond to chemical physics aspects of atoms in molecules depending on how
the model's parameters are penalized and optimized. Their advantage over
human-engineered features is that they facilitate the flexibility of the
models, and when extracted with posterior inference they generalize well.
Nevertheless, latent features tend to be difficult to interpret and posterior
inference relies on Bayesian variational methods that are challenging to
train\cite{Fu2019}. The \lstinline{atomistic.features} module in \textsf{ML4Chem} 
provides
a class to ease latent feature extraction to train any of the available
atomistic ML algorithms. It uses AE algorithms, like the ones described in
Sections~\ref{sec:unsupervisedlearning} and~\ref{sec:hybrid}, and convert raw
features into latent variables by forward-propagating them through the
encoder part of the AE architecture. Listing~\ref{lst:latent} shows an
example of the \lstinline{LatentFeatures} featurization module where two
keyword arguments are passed to instantiate this class:

\begin{itemize}
  \item A tuple called \lstinline{args} contains the name of the type of raw
  features and a dictionary with their respective parameters to be computed
  and subsequently converted into latent variables with a trained AE.
  \item We also assign to a variable \lstinline{encoder} a dictionary with
  keys \lstinline{"model"}, and \lstinline{"params"} containing the paths
  on disk to load the model and its parameters.
\end{itemize}

After instantiation, latent variables are computed with the
\lstinline{.calculate()} class method as it was done with the
\lstinline{Gaussian} class, and are returned in the right structure to be
used as input to train other ML algorithms. Note that they can also be
converted to a Pandas data frame.

\clearpage
\begin{lstlisting}[caption={Extraction of latent features using the
\lstinline{atomistic.features} module of \textsf{ML4Chem}.},
label={lst:latent},language=Python]
from ml4chem.atomistic.features import LatentFeatures

ae_path = "my_autoencoder/"

# Arguments to build raw features
normalized = True
preprocessor = ("MinMaxScaler", {"feature_range": (-1, 1)})
args = (
    "Gaussian",
    {
        "preprocessor": preprocessor,
        "cutoff": 6.5,
        "normalized": normalized,
        "save_preprocessor": "iso.scaler",
        "overwrite": False,
    },
)
# Dictionary to load trained autoencoder
encoder = {"model": ae_path + "vae.ml4c", 
           "params": ae_path + "vae.params"}

features = LatentFeatures(encoder=encoder, features=args)
latent = features.calculate(
    inputs, 
    purpose="training", 
    data=data_handler, 
    svm=False
)
df = features.to_pandas()
\end{lstlisting}

\subsection{Models}

This section describes atomistic ML regression algorithms in \textsf{ML4Chem}
under the \lstinline{atomistic.models} module. At this point, it is important
to differentiate ML algorithms from ML models. ML algorithms refer to all
procedures and steps carried out to solve a determined ML task while models
are well-defined results of algorithms. Another important difference is that
ML models are fed inputs to infer or predict some output. Atomistic ML
algorithms exploit the physical phenomenon of ``locality'' where atoms are
fundamental entities and whose ML features can be extracted by measuring
interactions between each atom and its nearest-neighbor atoms. Predictions
$P$ are therefore calculated as the sum of many individual contributions as
shown in Eq.~\ref{eq:atomistic}:

\begin{equation}
    \label{eq:atomistic}
    P = \sum_{i=1}^{n} p_i (\mathbf{F}_i(\mathbf{R}_i)),
\end{equation}

where a local contribution $p_i$ is a functional of a feature mapping
function $\mathbf{F}_i$ that takes as arguments atom positions
$\mathbf{R}_i$, and chemical symbols. There are two possible flavors of these
algorithms: \textit{i)} a sub regression model for each chemical symbol in
the data set exists or \textit{ii)} a unique regression model is used for all
chemical element types. 

To implement new deep learning atomistic algorithms in \textsf{ML4Chem},
developers have to derive their classes by inheriting the structure from the
\lstinline{DeepLearningModel} base class shipped in the \lstinline{base.py}
file and shown in Listing~\ref{lst:deeplearningmodels}:

\begin{lstlisting}[floatplacement=h, caption={Abstract base class for
the implementation of new atomistic deep learning models under
\lstinline{atomistic.models} module of
\textsf{ML4Chem}.}, label={lst:deeplearningmodels},language=Python]
from abc import ABC, abstractmethod
import torch


class DeepLearningModel(ABC, torch.nn.Module):
    @abstractmethod
    def name(cls):
        """Return name of the class"""
        return cls.NAME

    @abstractmethod
    def __init__(self, **kwargs):
        """Arguments needed to instantiate the model"""
        pass

    @abstractmethod
    def prepare_model(self, **kwargs):
        """Prepare model for training or inference"""
        pass

    @abstractmethod
    def forward(self, X):
        """Forward propagation pass"""
        pass
\end{lstlisting}

Deep learning classes require a \lstinline{name()} method that returns the
name of the model, list of keyword arguments to instantiate the model using
the reserved \lstinline{__init__()} constructor, a
\lstinline{prepare_model()} method where parameters are initialized and the
algorithm is prepared for the purposes of training or inference. Finally, a
\lstinline{forward()} method is responsible to perform a forward pass and
return predictions. Support vector machine algorithms, require all these
methods except for the \lstinline{forward()} function. The fulfillment of
this structure enables inter-operability within \textsf{ML4Chem}.

\subsubsection{Supervised Learning}

Supervised learning refers to the ML task of determining a complex function
that maps inputs into outputs from labeled pairs of input/target examples.
Its application assumes there exists a good understanding of the structure
and existent classes of the data. It is worth noting that the interpretation
of these models is usually easier than the ones obtained from unsupervised
learning ML tasks.

In atomistic ML algorithms, the inputs correspond to attributes of molecules
such as atom positions, atom types, total charge, band-gap, etc. All these
can be directly used as features, or mapped applying domain knowledge rules. To
train the models, targets can be any scalar or vector quantity associated
with a molecule like total energy, dipole moment, or atomic forces.
Supervised algorithms tend to require large amounts of data, and
featurization tends to be biased because it is human-engineered. Thus, models
are usually prone to perform poorly beyond training set regimes. Also, the
data sets have to be designed with enough diversity to capture meaningful
underlying structures from input data. Active learning
protocols\cite{ricci2011introduction, Tang2018} are known to help in assuring
this diversity. After the training procedure, and assessment of the
predictive power of models by cross-validation\cite{geisser2017predictive} or
other model validation techniques, their parameters can be stored to perform
inference in unknown data.

In the following sections, we discuss the type of neural network
architectures and support vector machine algorithms supported in 
the \lstinline{atomistic.models} module of \textsf{ML4Chem}.

\paragraph{Neural Networks}\hfill

Neural Networks (NN) are algorithms inspired by how the human brain works.
Their building blocks are constituted by hidden layers with interconnected
neurons. NN can approximate any function with arbitrary accuracy and their 
outputs can be represented by the following equation:

\begin{equation}
    \mathbf{y} = \mathbf{WX} + \mathbf{b},
    \label{eq:linearnn}
\end{equation}

where $\mathbf{X}$ are the inputs, $\mathbf{W}$ are learnable parameters,
$\mathbf{b}$ are the biases and $\mathbf{y}$ are outputs.
Eq.~\ref{eq:linearnn} is nothing but a linear function and without any
extra modification, it can only fit data having a linear response. In
the context of deep learning, the product $\mathbf{WX}$ is a nested
function\cite{Burkov2019} composed of $l$ hidden layers that returns either a
vector or a scalar:

\begin{equation}
    \mathbf{y} = \mathbf{Wx} + \mathbf{b} = \mathbf{f}_4(\mathbf{f}_3(\mathbf{f}_2(\mathbf{f}_1(\mathbf{x})))).
    \label{eq:nestedexample}
\end{equation}

From the equation above, $l = 4$ means the model has four layers that output
a vector. The nested function $\mathbf{f}_l(\mathbf{z})$ is defined as:

\begin{equation}
    \mathbf{f}_l(\mathbf{z}) = \mathbf{a}_l(\mathbf{W}_l\mathbf{z} + \mathbf{b}_l),
    \label{eq:nested}
\end{equation}

where we have introduced an \textit{activation function} denoted by
$\mathbf{a}_l$. The effect of the activation function on the outputs
$\mathbf{z}$ of the neurons is to add non-linear response making the NN
suitable to approximate more complex non linear functions. Some of the most
common activation functions are $tanh(z)$, $sigmoid(z)$, and $relu(z)$:

\begin{equation}
    tanh(z) = \frac{(e^z - e^{-z})}{(e^z + e^{-z})}
    \label{eq:tanh}
\end{equation}

\begin{equation}
    \sigma(z) = \frac{1}{1 + e^{-z}}
    \label{eq:sigmoid}
\end{equation}

\begin{equation}
    relu(z) =
    \begin{cases}
        0 &\text{for $z \leq 0$}\\
        z &\text{for $z \geq 0$}
    \end{cases}
    \label{eq:relu}
\end{equation}

When forward-propagating information through NN, the type of activation
function applied to the last layer determines whether the task is 
regression (linear or non-linear activation function), or classification
(logistic activation function)\cite{Goodfellow2016}. When more than
two hidden layers are stacked between the input and output layers, NN are
called \textit{deep neural networks}.

\begin{figure}[!ht]
    \centering
    \includegraphics[width=12cm]{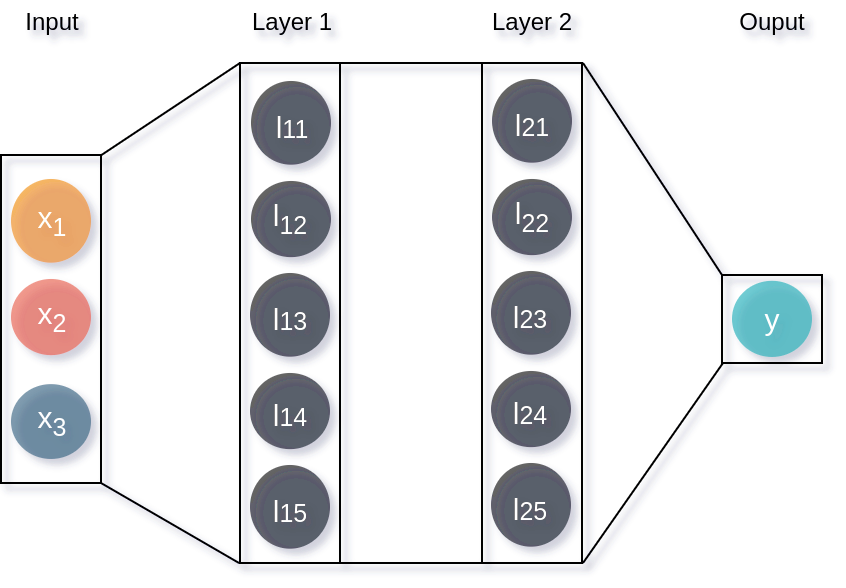}
    \caption{Schematic representation of a neural network.}
    \label{fig:nn}
\end{figure}

In Figure~\ref{fig:nn} we show a neural network with one input layer, two
hidden layers, and an output layer. In this multi-layer perceptron, we assume
all outputs of a layer are connected to the inputs of succeeding layers.

Our atomistic neural network implementation follows the structure described
by Behler and Parrinello\cite{Behler2007} where each set of chemical elements
in the data set has its own NN algorithm. For instance, if the data set
contains C, H, N, and O atoms, then there will be four different NN. Hidden
layers are \textit{fully connected} and the same activation function is
applied to each of them. Different activation functions to each hidden layer
and application of convolutions are features planned for future releases.
However, various NN with different activation functions can be merged and
trained simultaneously with the \lstinline{ModelMerger} module described in
Section~\ref{sec:hybrid}. NN algorithms are trained by forward-propagating
the atomistic feature vectors of molecules through their respective NN. The
outputs of the models are atomic contributions. If targets are global
quantities, such as the total energy, then the local atomic energy
contributions of a molecule can be summed up and used to evaluate the loss
function against corresponding targets. In this way, when
backward-propagating the models, the optimization process will account for
the global quantity. The usage of NN in the \lstinline{atomistic.models}
module is illustrated in Listing~\ref{lst:nn}. A \lstinline{NeuralNetwork}
class with two hidden layers of 10 nodes each is instantiated, and a ReLU
activation function is applied to all hidden layers except for the output
layer. The model is subsequently prepared for training with atomistic feature
vectors holding 4 dimensions per atom. Note that we also feed the
\lstinline{data_handler} object created with the \lstinline{Data} class.

\begin{lstlisting}[caption={Usage example script of the atomistic neural
network class in \textsf{ML4Chem}.}, label={lst:nn},language=Python]
from ml4chem.atomistic.models import NeuralNetwork

# Parameters
n = 10                  # Hidden layers
activation = "relu"     # Activation function
input_dimension = 4     # Input Dimension

# Model instantiation and preparation
nn = NeuralNetwork(hiddenlayers=(n, n), activation=activation)
nn.prepare_model(
    input_dimension, 
    data=data_handler, 
    purpose="training"
)
\end{lstlisting}

\paragraph{Support Vector Machines}\hfill

A Support Vector Machine (SVM) is a discriminative classifier that for some
given labeled examples, it outputs an optimal hyperplane which categorizes
new examples. For atomistic ML models, Kernel Ridge Regression (KRR) is a
very useful algorithm based on linear ridge regression that aims to minimize
a squared error loss function with a $l2$-norm regularization term:

\begin{equation}
 \label{eq:ridge}
 \operatorname*{argmin}_{\mathbf{\beta} \in \mathbb{R}^D} \sum_{i=1}^N
 (\mathbf{\beta}\mathbf{x_i} - y_i)^2 + \lambda ||\mathbf{\beta}||^2_2,
\end{equation}

where $\mathbf{\beta}$ are regression coefficients, $\mathbf{x_i}$ is the $i^{th}$ model
input, $y_i$ is its corresponding $i^{th}$ target, and $\lambda \geq 0$ is a
hyperparameter representing the penalty during the minimization of the loss
function.
Although the loss function in Eq.~\ref{eq:ridge} is only suitable for linear
regression, it can be applied to non-linear problems using the \textit{kernel
trick} (KT)\cite{P.Murphy2012}. The KT allows model inputs (atom features in
this framework) to be mapped into other feature spaces through the use of
kernel functions (KF). The importance of KF lies in that they are computed
directly in raw input space (atomistic feature vectors are not explicitly
modified). KF are non-negative real-valued functions between two vectors
$k(\mathbf{x}, \mathbf{y})$. According to \textit{Mercer's theorem}, matrices
built from KFs are squared definite-positive covariant kernel
matrices\cite{Marsland2015}. 
Thus, they provide some numerical advantages
because they can be factorized and generally provide well-defined solutions.
The KRR algorithm of the atomistic module of \textsf{ML4Chem} supports radial basis kernel functions (RBF), squared
exponential kernel, linear, and Laplacian kernels in both their isotropic and
anisotropic\cite{Sowmya2004} variants as seen on Eqs.~\ref{eq:rbfiso} and
\ref{eq:rbfaniso} respectively:

\begin{equation}
 \label{eq:rbfiso}
 k(\mathbf{x},\mathbf{y})_{rbf}^{iso} = \exp{\left(-\frac{||\mathbf{x}
 - \mathbf{y}||^2_2}{2\sigma^2}\right)},
\end{equation}

\begin{equation}
 \label{eq:rbfaniso}
 k(\mathbf{x},\mathbf{y})_{rbf}^{aniso} = \exp{\left(-\sum_{i=1}^D\frac{(x_i
 - y_i)}{2\sigma^2_i}\right)},
\end{equation}

where $\mathbf{x}$ and $\mathbf{y}$ are input and reference atom feature
vectors respectively, $\sigma$ is the Gaussian width, and $D$ is the number
of dimensions of the feature vectors. These kernels are bivariate functions
that compute the similarity based on distances between two vectors in a
normalized vector space. The output of these KF is within the interval $[0, 1] $.
Note that anisotropic variants of kernels allow the assignment of specific
variances for each dimension of the feature vectors without the need
to modifying them explicitly.

Our implementation is inspired by Ref.~\cite{Rupp2015} and uses the atomistic
feature vectors to build the kernel matrix shown in
Eqs.~\ref{eq:kernel_energy}. 
Kernel matrices are positive definite which make their Cholesky decomposition
to provide a unique solution to the regression coefficients. This is a
particular advantage of these models since their solution is deterministic.
By deterministic we mean a model that for a given input always reproduces the
same regression coefficients, and consequently the same output. That is not
the case in models such as neural networks that are randomly initialized
leading to different local minima that are valid solutions.

\begin{equation}
 \label{eq:kernel_energy}
\begin{bmatrix}
 k_{11}^E & k_{12}^E & k_{13}^E & \dots & k^E_{1N} \\
 k_{21}^E & k_{22}^E & k_{23}^E & \dots & k^E_{2N} \\
 \hdotsfor{5} \\
 k_{N1}^E & k_{N2}^E & k_{N3}^E & \dots & k^E_{NN}
\end{bmatrix}
\begin{bmatrix}
 \beta_{1} \\
 \beta_{2} \\
 \vdots \\
 \beta_{N} \\
\end{bmatrix}
=
\begin{bmatrix}
 E_{1} \\
 E_{2} \\
 \vdots \\
 E_{N} \\
\end{bmatrix}
\end{equation}

Solving the system of equations in Eq.~\ref{eq:kernel_energy} requires atomic
quantities that might not be available in quantum mechanics such as atomic
energies. In those cases, one has to rely on atomic decomposition Ansatz as
described by Bart\'ok\cite{Bartok2010, Bartok2013, Bartok2015}. Instead of
training using a loss function, the regression coefficients $\beta$ in
Eq.~\ref{eq:kernel_energy} are determined by forward and backward 
substitution after matrix factorization using the Cholesky
decomposition method.

After training, a prediction is carried out by

\begin{equation}
 f(\mathbf{x}) = \sum_{i=1}^{n}\beta_i k(\mathbf{x_i},\mathbf{y_j}).
\end{equation}

The \lstinline{atomistic} module of \textsf{ML4Chem} can perform kernel ridge
regression as shown in Listing~\ref{lst:krr}. The \lstinline{KernelRidge}
class is instantiated with the kernel function to be used, the isotropic
variance \lstinline{sigma} and penalization values. Then, the model is
prepared by passing the feature and reference features to build the
covariance matrix. We also support Gaussian process regression in which case
the listing has to be modified to import the \lstinline{GaussianProcess}
class instead. When doing so, the predictions will also return a tuple with
the associated predictive uncertainty.

\clearpage
\begin{lstlisting}[caption={Usage example script of the atomistic Kernel
ridge regression class in ML4Chem.}, label={lst:krr},language=Python]
from ml4chem.atomistic.models import KernelRidge

# Parameters
sigma = 1.0             # Sigma kernel value
kernel = "rbf"          # Kernel type
lamda = 1e-5            # Penalization

# Model instantiation and preparation
krr = KernelRidge(sigma=sigma, kernel=kernel, lamda=lamda)
krr.prepare_model(
    feature_space, 
    reference_features, 
    data=data_handler
)
\end{lstlisting}

\subsubsection{Unsupervised Learning}
\label{sec:unsupervisedlearning}

\begin{figure}[ht]
    \centering
    \includegraphics[width=8cm]{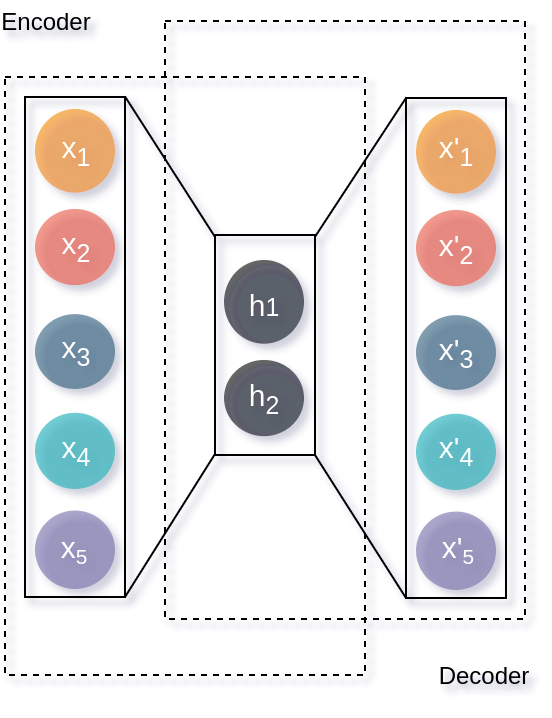}
    \caption{Architecture of an Autoencoder.}
    \label{fig:ae}
\end{figure}

Probably the best example of an unsupervised learning algorithm corresponds
to autoencoders (AE). AE are a type of artificial neural network
architecture composed by an encoder and a decoder that can learn data
representations without human intervention. As shown in Figure~\ref{fig:ae},
an AE forward propagates the input through the encoder architecture to reach
an informational bottleneck where latent features, denoted with $h_d$, are
extracted. The bottleneck layer is usually of lower dimensionality compared
to the input. Afterward, these latent features are used by the decoder to
reconstruct the inputs. This reconstruction task might seem uninteresting but
depending on the type of AE architecture, and what is being reconstructed,
the model can learn low-dimensional representations of the input space,
denoise data, or even how to generate new examples.

An autoencoder with fully connected hidden layers tends to only ``memorize''
how to reconstruct the input space or denoise data. Their power can be enhanced by
penalizing the latent space with a loss
function\cite{Lemke2019}, attachment of external
task\cite{Gomez-Bombarelli2018}, or just by forcing sparse
activations\cite{Makhzani2013,Arpit2015}.

Our implementation of autoencoders supports two architectures that can be set with
the \lstinline{one_for_all} boolean keyword argument. If set to
true, a single autoencoder is used for all types of atoms in the training
data, otherwise, each set of atom type will have its autoencoder as in the
Behler-Parrinelo scheme. In Listing~\ref{lst:aeml4chem}, an
\lstinline{AutoEncoder()} class in the \lstinline{atomistic.models} module of
\textsf{ML4Chem} is instantiated with an encoder/decoder architecture where
input's dimensions are lowered from 40 to 4 dimensions. In this case, the hyperbolic
tangent activation function is applied to all layers
with the \lstinline{activation} keyword argument, and when the \lstinline{purpose} is not set then training is assumed. Currently, only the same activation function for all
layers is supported but this functionality can be extended in future
releases. Note that the \lstinline{.prepare_model()} class method sets the
dimensionality of the output to be the same as the input.

\begin{lstlisting}[float, floatplacement=H, caption={Training example script
of the atomistic autoencoeder class in ML4Chem.},
label={lst:aeml4chem},language=Python]
from ml4chem.atomistic.models import AutoEncoder


hiddenlayers = {"encoder": (20, 10, 4), "decoder": (4, 10, 20)}
activation = "tanh"
purpose = "training"
input_dimension = 40

ae = AutoEncoder(
    hiddenlayers=hiddenlayers, 
    activation=activation, 
    one_for_all=False
)
ae.prepare_model(input_dimension, input_dimension, data=data_handler)
\end{lstlisting}

\begin{figure}[!ht]
    \centering
    \includegraphics[width=12cm]{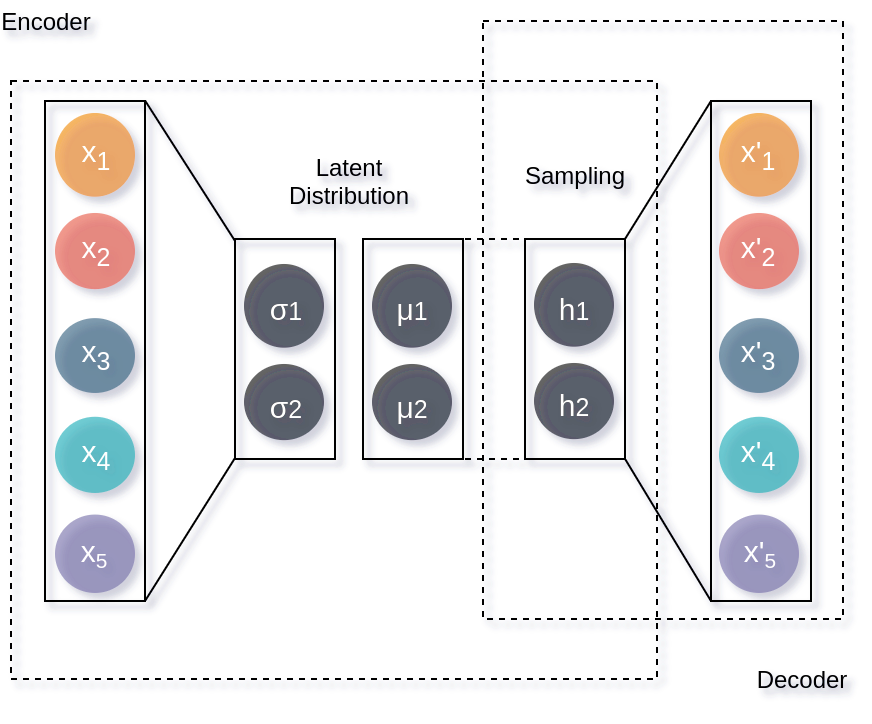}
    \caption{Variational Autoencoder architecture.}
    \label{fig:vae}
\end{figure}

In 2013 Kingma and Welling \cite{Kingma2013} proposed a modification to
autoencoder architectures employing a variational Bayes method that infers a
posterior probability. Instead of encoding a latent vector, as seen in
Figure~\ref{fig:vae}, the variational autoencoder (VAE) encodes a
non-differentiable latent normal distribution with unit variance. To make the
model trainable, a latent vector is randomly sampled from the distribution
and used by the decoder to reconstruct the input space. This is known as the
\textit{reparameterization trick} and allows the calculation of the gradient
of the loss function with respect to the parameters of this
architecture\cite{Kingma2013}. VAEs are known to produce blurry
reconstructions\cite{7926714,Zhao2017} of the inputs, but more meaningful
data representations, unlike vanilla AEs. That is because the VAEs are forced
to minimize reconstruction errors from feature vectors sampled from the
latent distribution. VAEs are classified as \textit{generative models}
because they learn smooth conditional distributions of the input space
$\mathbf{X}$ for given evidence and are even able to generate new examples.
VAEs can be invoked in \textsf{ML4Chem} by importing \lstinline{VAE()}
instead of the \lstinline{AutoEncoder()} class in
Listing~\ref{lst:aeml4chem}.

In our VAE implementation, we also support the \lstinline{one_for_all}
keyword argument, and provide three variants when passing the
\lstinline{variant} string keyword argument:

\begin{enumerate}
\item "multivariate": the decoder outputs a distribution with mean and
unit variance, and the model is trained by minimizing the negative of the
log-likelihood plus the Kullback–Leibler
divergence\cite{kullback1951information}. This is useful when outputs are
continuous variables. Expected feature range $[-\infty, \infty]$.
\item "bernoulli": the sigmoid activation function is applied to the decoder's
outputs and models are also minimized with the negative of the
log-likelihood plus the Kullback–Leibler
divergence\cite{kullback1951information}. Features must be in a range $[0, 1]$.
\item "dcgan": hyperbolic tangent activation function is applied to the
decoder's outputs, and the model is trained by minimizing  the negative of the
log-likelihood plus the Kullback–Leibler
divergence\cite{kullback1951information}. Useful for feature ranges $[-1, 1]$.
\end{enumerate} 

\subsubsection{Semi-Supervised and Hybrid Learning}
\label{sec:hybrid}

In previous sections, we have discussed the supervised and unsupervised
learning algorithms. Supervised learning requires large amounts of data to be
labeled in input/target pairs which usually requires a very costly process
carried out by an ML engineer or a data scientist. On the other hand,
unsupervised learning does not require labeled data but its applicability is
very limited and dependent on the problem that is tried to be solved. 
To overcome these limitations, semi-supervised learning tasks consist of
building algorithms that can work with both labeled and unlabeled data.
Mixing both tasks is challenging as several loss functions are used to then 
backward propagate the ensemble model.

The \textsf{atomistic.models} module allows access to semi-supervised
learning with the \lstinline{ModelMerger()} class. In
Listing~\ref{lst:ssml4chem}, we use the neural network and autoencoder models
already instantiated in Listings \ref{lst:nn} and \ref{lst:aeml4chem} and add
them to a list called \lstinline{models}. We import the required loss
functions from the \lstinline{models.loss} module and assigned them to a list
named \lstinline{losses}. Similarly, the targets for each model are added to
a list with name \lstinline{targets}, and for the inputs, note that the neural
network model takes as input the latent space from the autoencoder.

During training, there are two important options that can be
passed to the \lstinline{ModelMerger} class: 

\begin{itemize}
 \item A boolean \lstinline{"independent_loss"} keyword argument to set 
 whether or not the loss functions are merged. If set to true, 
 models are aware of each other. 
 \item A \lstinline{"loss_weights"} keyword argument with a list to set how
 much the loss of model(i) contributes to the total loss function.
\end{itemize}

More about training procedures is elaborated in Section~\ref{sec:modelopt}.

\begin{lstlisting}[caption={Example of a hybrid
model in \textsf{ML4Chem} combining an autoencoder with a neural network.},
label={lst:ssml4chem},language=Python]
from ml4chem.atomistic.models.loss import MSELoss, AtomicMSELoss

models = [ae, nn]
losses = [MSELoss, AtomicMSELoss]

# AE input/output dimensions are 20/4.
# NN input/output dimensions are 4/1.
inputs = [X, ae.get_latent_space]

# AE targets are X. 
# NN targets are vector of scalars y.
targets = [X, y]

merged = ModelMerger(models)
merged.train(
        inputs=inputs,
        targets=targets,
        data=data_handler,
        lossfxn=losses
)
\end{lstlisting}

\clearpage
\subsection{Model Optimization}
\label{sec:modelopt}

Deep learning is a challenging optimization problem that employs the gradient
descent algorithm. Partial derivatives of the loss function with respect to
the model's parameter space are required to update ML model's parameters in
the direction of steepest descent as defined by the negative of the gradient.
This computation experiences the vanishing or exploding gradient problems. In
the former case, the partial derivatives of the loss function with respect to
the parameters of the model become so small that the optimizers cannot adjust
weights to minimize or maximize the loss function. In the latter case,
partial derivatives are so large that optimizers oscillate around a minimum
or maximum in the loss function space and convergence is never reached. These
problems remained elusive for decades but were resolved by gradient
clipping\cite{pascanu2012understanding}, regularization\cite{Burkov2019}, and
usage of activation functions that only saturate in one direction
\textit{e.g.} the ReLU activation function\cite{glorot2011deep}. The
\lstinline{atomistic} module of \textsf{ML4Chem} supports most of the
optimizers available in Pytorch such as ADAM\cite{Kingma2014}, stochastic
gradient descend\cite{robbins1951stochastic}, and
Adagrad\cite{duchi2011adaptive}.

\begin{figure}[h]
    \centering
    \includegraphics[width=12cm]{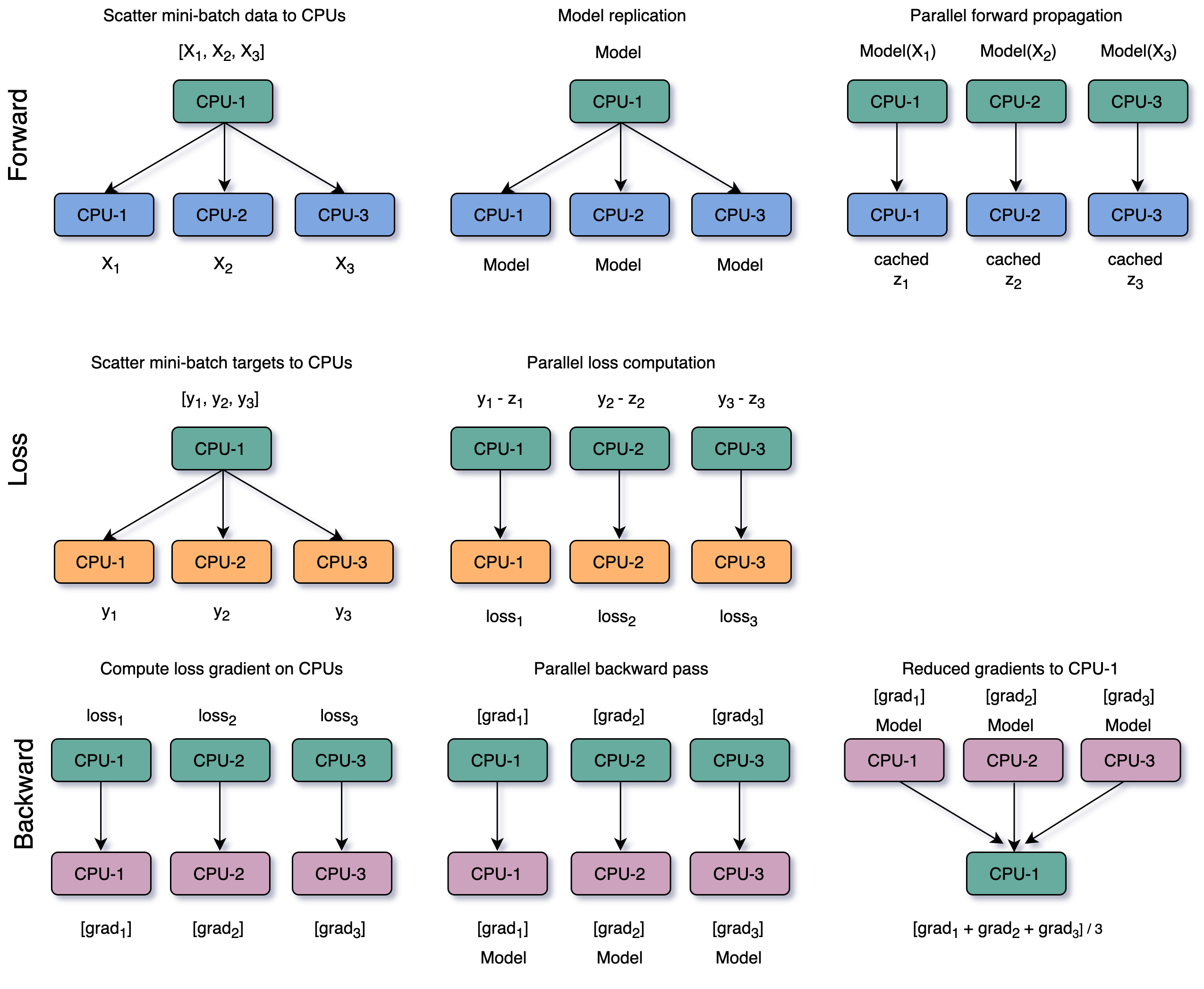}
    \caption{Mini-batch data parallel scheme.}
    \label{fig:dataparallel}
\end{figure}

\begin{figure}[ht]
    \centering
    \includegraphics[width=12cm]{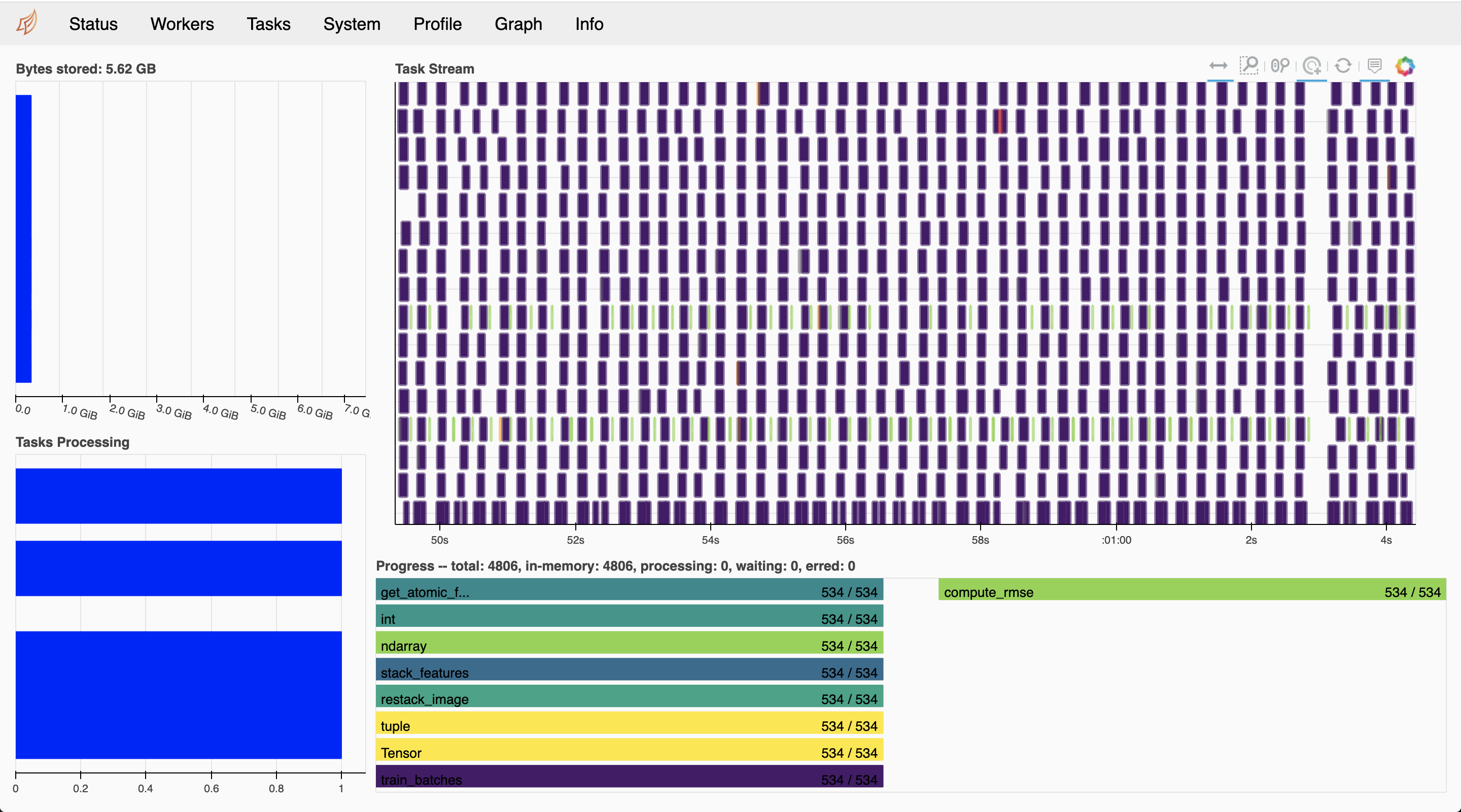}
    \caption{Dask dashboard tool tracking computations in real time for
             the case \\
             where \textsf{ML4Chem} is optimizing a neural network
             over 16 processes.}
    \label{fig:dask}
\end{figure}

We train models exploiting the data parallelism paradigm as illustrated in
Figure~\ref{fig:dataparallel}. In this scheme, data is partitioned in
mini-batches that are set with the \lstinline{batch_size} keyword argument.
Mini-batches with the input data and targets are scattered through workers in
a local or remote distributed cluster. The ML algorithm also is replicated on
each of these workers, and parallel forward propagation is carried out. After
the forward pass, we proceed with the parallel evaluation of the loss
function for each mini-batch. In this step, the Pytorch automatic
differentiation package \lstinline{autograd} computes the gradient of the
loss functions with respect to the weights of the model for each mini-batch
and performs parallel backward-propagation passes. Finally, the gradients are
reduced in CPU-1 and we call a step in the optimizer to update weights
according to this gradient. This process is repeated until the number of
epochs is exhausted or some convergence criterion is reached. The
parallelism scheme described above can also be visualized with the Dask
dashboard as shown in Figure~\ref{fig:dask}.

Each algorithm in the \lstinline{atomistic.models} module provides a
\lstinline{train} class to optimize its parameters. Each model has its
particularities but, in general, the \lstinline{train} class requires at
least input, outputs, batch size, optimizer, and loss function (if needed).
The \lstinline{train} class is responsible for taking all this information
and carry out all necessary steps to train a model including calling steps in
the optimizer.
The optimizers are set with the \lstinline{atomistic.models.optim} module
that wraps Pytorch optimizer objects from a tuple composed by the name of the
optimizer, and a dictionary with its required parameters \textit{e.g.}:
\lstinline|optimizer = ("adam", {"lr": float, "weight_decay"=float})|.
When ML models are trained, their parameters can be stored to disk and
subsequently, be used for prediction.

\subsection{Inference and Visualization}

Inference is the step in an ML pipeline where a trained ML model is used to
infer/predict over unseen samples. This procedure involves a similar forward
pass as the carried out during training for the prediction of targets.

Probably the biggest difference with a model in training mode is that during
inference no backward-propagation is carried out. Therefore, the inference is
a very fast computation that requires matrix multiplication and summation. In
\textsf{ML4Chem}, atomistic models can be saved and loaded as seen in
Listing~\ref{lst:inference}.

We use the \lstinline{.save()} method of the \lstinline{Atomistic} class to
save a neural network and gaussian features passing as arguments their
objects. The \lstinline{label} argument is used to save them to disk with
name \lstinline{"nn"}. A trajectory file is loaded in memory that contains
molecules stored as \lstinline{Atoms()} ASE objects. Then, we assign the
model to the \lstinline{calc} variable with the \lstinline{load()} function
and the following arguments: \textit{i)} path to optimized parameters stored
in a file named \lstinline{nn.ml4c}, \textit{ii)} path to the
\lstinline{nn.params} file that contains parameters in JSON format to
recreate the model and features, and \textit{iii)} the preprocessor to scale
the features from \lstinline{nn.scaler}.

\medskip
\begin{lstlisting}[caption={Example of save, and load trained neural network
model in \textsf{ML4Chem}.},
label={lst:inference},language=Python]
from ml4chem.atomistic import Atomistic
from ase.io import Trajectory

# Save a model
Atomistic.save(nn, features=gaussian, path="", label="nn")

# Load a model 
molecules = Trajectory("test.traj")
calc = Atomistic.load(model="nn.ml4c", params="nn.params", preprocessor="nn.scaler")    
     
for molecule in molecules:
    energy = calc.get_potential_energy(molecule)
\end{lstlisting}

\begin{figure}[h]
\begin{subfigure}[ht]{0.40\linewidth}
\includegraphics[width=\linewidth]{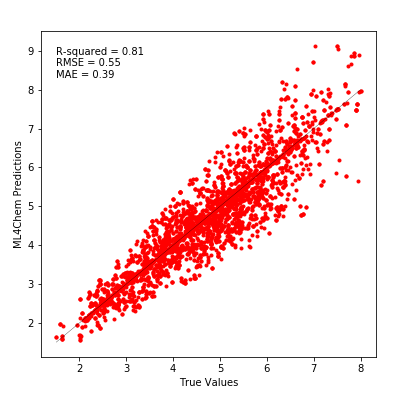}
\caption{Parity plot.}
\end{subfigure}
\hfill
\begin{subfigure}[ht]{0.52\linewidth}
\includegraphics[width=\linewidth]{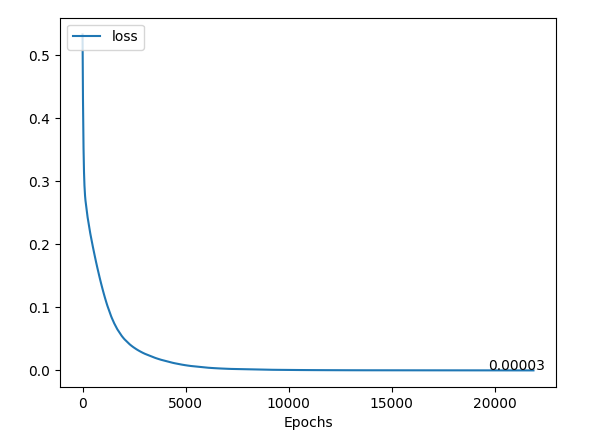}
\caption{Loss function progress.}
\end{subfigure}%

\begin{subfigure}[ht]{0.45\linewidth}
\includegraphics[width=\linewidth]{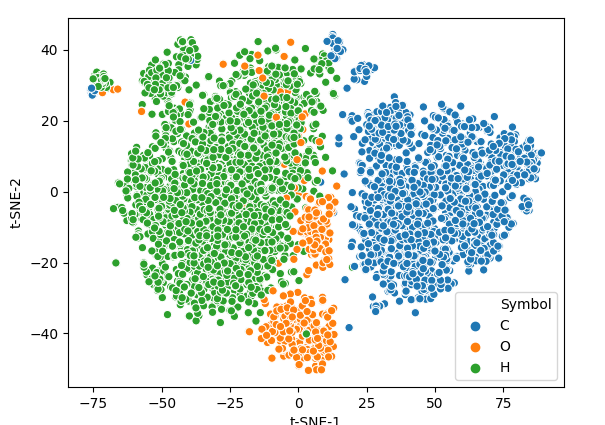}
\caption{2D latent space (\lstinline{seaborn}).}
\label{fig:2dlatent}
\end{subfigure}
\hfill
\begin{subfigure}[ht]{0.45\linewidth}
\includegraphics[width=\linewidth]{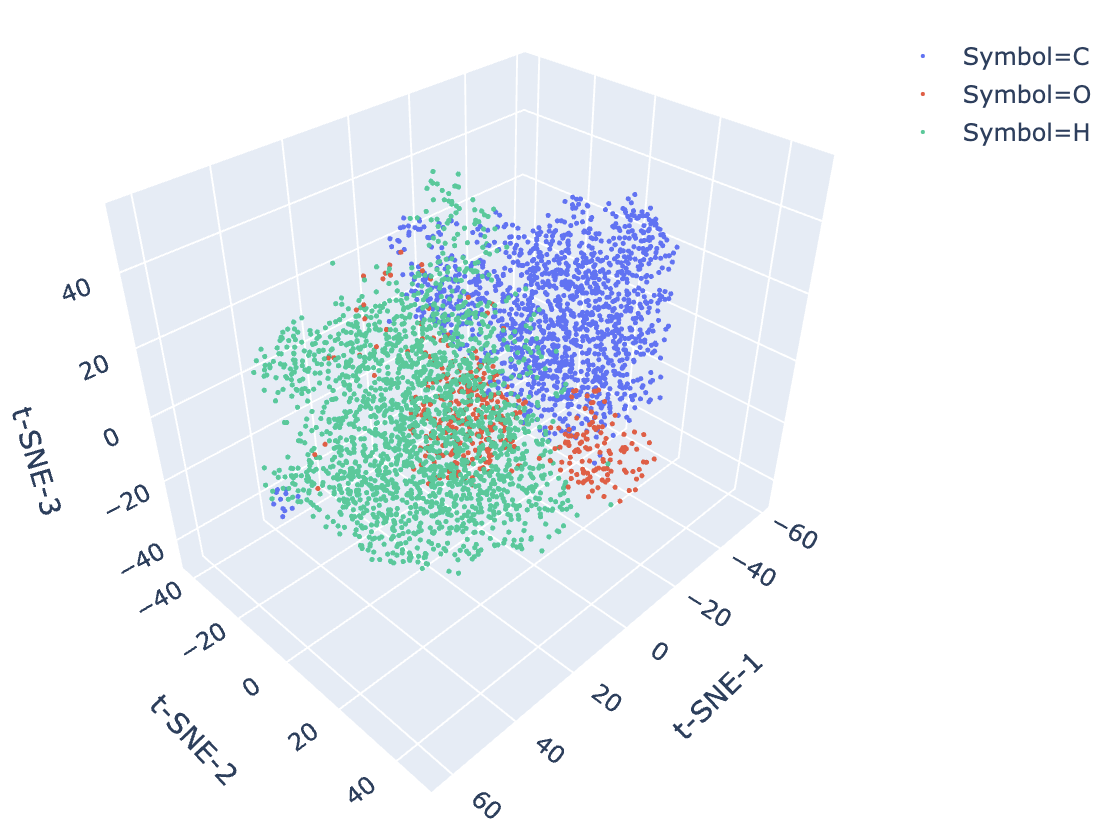}
\caption{3D latent space (\lstinline{plotly}).}
\end{subfigure}%
\caption{Visualization tools offered in \textsf{ML4Chem}.}
\label{fig:visualization}
\end{figure}

Another important part of ML pipelines has to do with visualization. We
provide a \textsf{visualization} module built on top of
matplotlib\cite{Hunter:2007}, seaborn and plotly\cite{plotly}. In
Figure~\ref{fig:visualization} we show a parity plot between \textsf{ML4Chem}
predictions and true values, the progress of the loss function in real-time,
a 2D visualization of atomistic latent space obtained using a VAE, and its 3D
interactive visualization with \lstinline{plotly}. All these plots are offered as
standalone functions. We also provide a command-line tool called
\textsf{ml4chem} that allows quick access to these visualizations. For
example, the latent space visualization shown in Figure~\ref{fig:2dlatent}
can be plotted from a stored features file using the command line 
\lstinline{ml4chem --plot tsne --file latent_space.db}.

\section{Demonstrations of the \textsf{ML4Chem} Framework}
\label{sec:demonstrations}

In this section, we present some demonstrations of running ML pipelines with
\textsf{ML4Chem}. An atomistic neural network is trained with the ANI-1 data set,
and a kernel ridge regression algorithm is trained with the QM7 data set.

\subsection{Data Sets}

The ANI-1 data set\cite{Smith2017} is publicly available at
\url{https://github.com/isayev/ANI1_dataset}. It is provided in HDF5 format,
and can be used with the \lstinline{atomistic} module of \textsf{ML4Chem}
when converted to ASE and passed to the \lstinline{Data} class. In
Listing~\ref{lst:anitoase}, we show how to load ANI-1 files to memory with \lstinline{pyanitools}  and assign
them to a \lstinline{ani_data} variable. Next, this list is passed as an
argument to the \textsf{ML4Chem}'s \lstinline{ani_to_ase()} parser function that converts HDF5
ANI-1 data sets to ASE trajectory files. After the conversion is done, the
trajectory file is saved to disk as \lstinline{ani.traj}.

\begin{lstlisting}[floatplacement=h, caption={Conversion of ANI-1 data set to the atomic simulation 
environment format (ASE).}, label={lst:anitoase},language=Python]
import ase
import pyanitools as pya
from ml4chem.data.parser import ani_to_ase
from ml4chem.data.utils import split_data

# Load ANI-1 hdf5 files to a list
files = ["ani_gdb_s01.h5", "ani_gdb_s02.h5","ani_gdb_s03.h5"]
ani_data = [pya.anidataloader(f) for f in files]

# Pass list of hdf5 files to ML4Chem ani_to_ase() parser function.
ani_dataset = ani_to_ase(
    ani_data, 
    data_keys=["energies"], 
    trajfile="ani.traj"
)

# Load trajectory file and split data in 80% training and 20% test set
traj = ase.io.Trajectory("ani.traj")
split_data(traj, test_set=20, randomize=True)
\end{lstlisting}

The QM7\cite{rupp2012data, blum2009} is available at
\url{http://quantum-machine.org/datasets} as a Matlab file. According to the
website, ``The data set is composed of three multidimensional arrays X (7165
x 23 x 23), T (7165) and P (5 x 1433) representing the inputs (Coulomb
matrices), the labels (atomization energies) and the splits for
cross-validation, respectively. The data set also contains two additional
multidimensional arrays Z (7165) and R (7165 x 3) representing the atomic
charge and the Cartesian coordinate of each atom in the molecules''. Its
conversion into an ASE trajectory file to be used in \lstinline{atomistic}
module of \textsf{ML4Chem} is trivial as illustrated in Listing~\ref{lst:qm7}. 

\begin{lstlisting}[caption={Conversion of QM7 data set to the atomic simulation 
environment format (ASE).}, label={lst:qm7},language=Python]
import scipy.io as sio
import ase
from ml4chem.data.utils import split_data


traj = ase.io.Trajectory("qm7.traj", mode="w")
dataset = sio.loadmat("qm7.mat")

# Cartesian coordinates
cartesian = dataset["R"]

# Atomic charge (Z), atomization energy (T)
Z = dataset["Z"]
T = dataset["T"]

for i, molecule in enumerate(Z):
    numbers, positions = [], []
    for j, z in enumerate(molecule):
        if z != 0:
            numbers.append(z)
            pos = tuple(cartesian[i][j])
            pos = [c * ase.units.Bohr for c in pos]
            positions.append(pos)
    atoms = ase.Atoms(numbers=numbers, positions=positions)

    traj.write(atoms, energy=float(T[0][i]))

# Load trajectory file and split data in 80% training and 20% test set
traj = ase.io.Trajectory("qm7.traj")
split_data(traj)
\end{lstlisting}

Each data set was randomized and split in $80\%$ as training set and $20\%$
as test set using the \lstinline{split_data()} function.

\subsection{Neural Network with Gaussian Features}

In this demonstration we trained an atomistic machine learning potential
using the \lstinline{NeuralNetwork} class with Gaussian type features and the
ANI-1 data set. For this example, we will use the high-level
\lstinline{Potentials} class. 

\begin{lstlisting}[caption={Training a neural
network algorithm in \textsf{ML4Chem} using the \lstinline{Potentials}
class.}, label={lst:nntraining},language=Python]
from ase.io import read
from ml4chem.utils import logger
from ml4chem.atomistic.features import Gaussian
from ml4chem.atomistic.models import NeuralNetwork
from ml4chem.atomistic import Potentials
from dask.distributed import Client, LocalCluster


def run():
    # Use 500 molecules for a total of 3606 feature vectors
    n_molecules, batch_size = 500, 30
    logger("nn.log")

    training = read("../training_images.traj", index="0:{}:1".format(n_molecules))

    # Instatiate the Potentials class 
    calc = Potentials(
        features=Gaussian(
            batch_size=batch_size, cutoff=6.5, normalized=True, save_preprocessor="model.scaler"
        ),
        model=NeuralNetwork(
            hiddenlayers=(10, 10), activation="tanh"
        ),
        label="nn_training",
    )

    # Optimizer options and convergence criterion
    convergence = {"energy": 5e-2}
    lr = 1.0e-2
    weight_decay = 1e-5
    optimizer = ("adam", {"lr": lr, "weight_decay": weight_decay})

    # Train the algorithm
    calc.train(
        training_set=training, convergence=convergence, optimizer=optimizer, batch_size=batch_size
    )


if __name__ == "__main__":
    cluster = LocalCluster(n_workers=16, threads_per_worker=1)
    client = Client(cluster)
    run()
\end{lstlisting}

In Listing~\ref{lst:nntraining} the
\lstinline{Potentials} class can be instantiated with any of the
\lstinline{atomistic.features} and \lstinline{atomistic.models} objects of
\textsf{ML4Chem}, and automates all of the necessary steps to train an
atomistic ML potential.

The architecture of the NN is of two hidden layers of 10 nodes each, a
hyperbolic tangent activation function, and the ADAM optimizer. The
convergence criterion is set to be a root-mean-squared error of
$0.05$~Hartree. The optimization converged at about 5000 epochs.

\begin{figure}[ht]
    \centering
    \includegraphics[width=12cm]{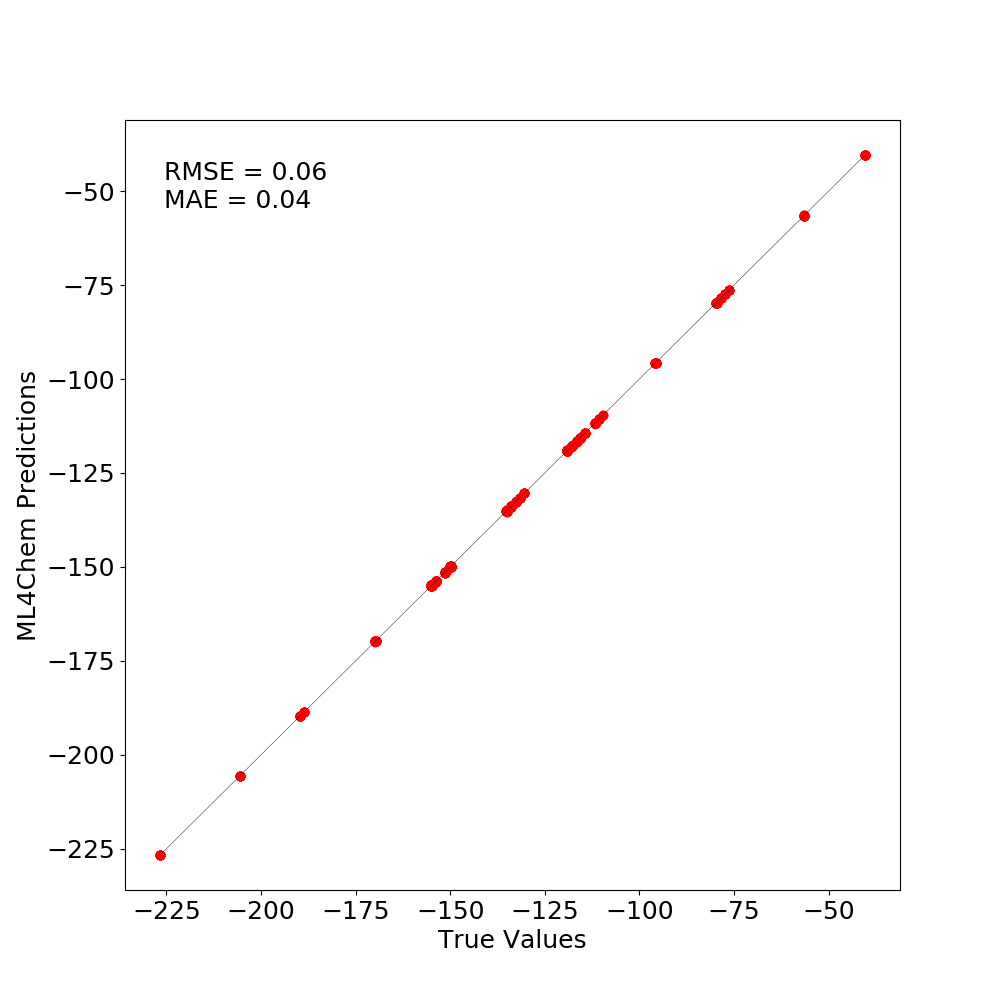}
    \caption{Parity plot of neural network atomistic model predictions over 
    $1000$ \\ molecules in the ANI-1 data set. Units are in Hartree.}
    \label{fig:nngaussianANI}
\end{figure}

In Figure~\ref{fig:nngaussianANI} we show a parity plot with results obtained
by predicting over 1000 unknown molecules of the test set. The NN model can
do predictions with high accuracy, achieving a MAE $0.04$~Hartree. 

\begin{figure}[ht]
    \centering
    \includegraphics[width=12cm]{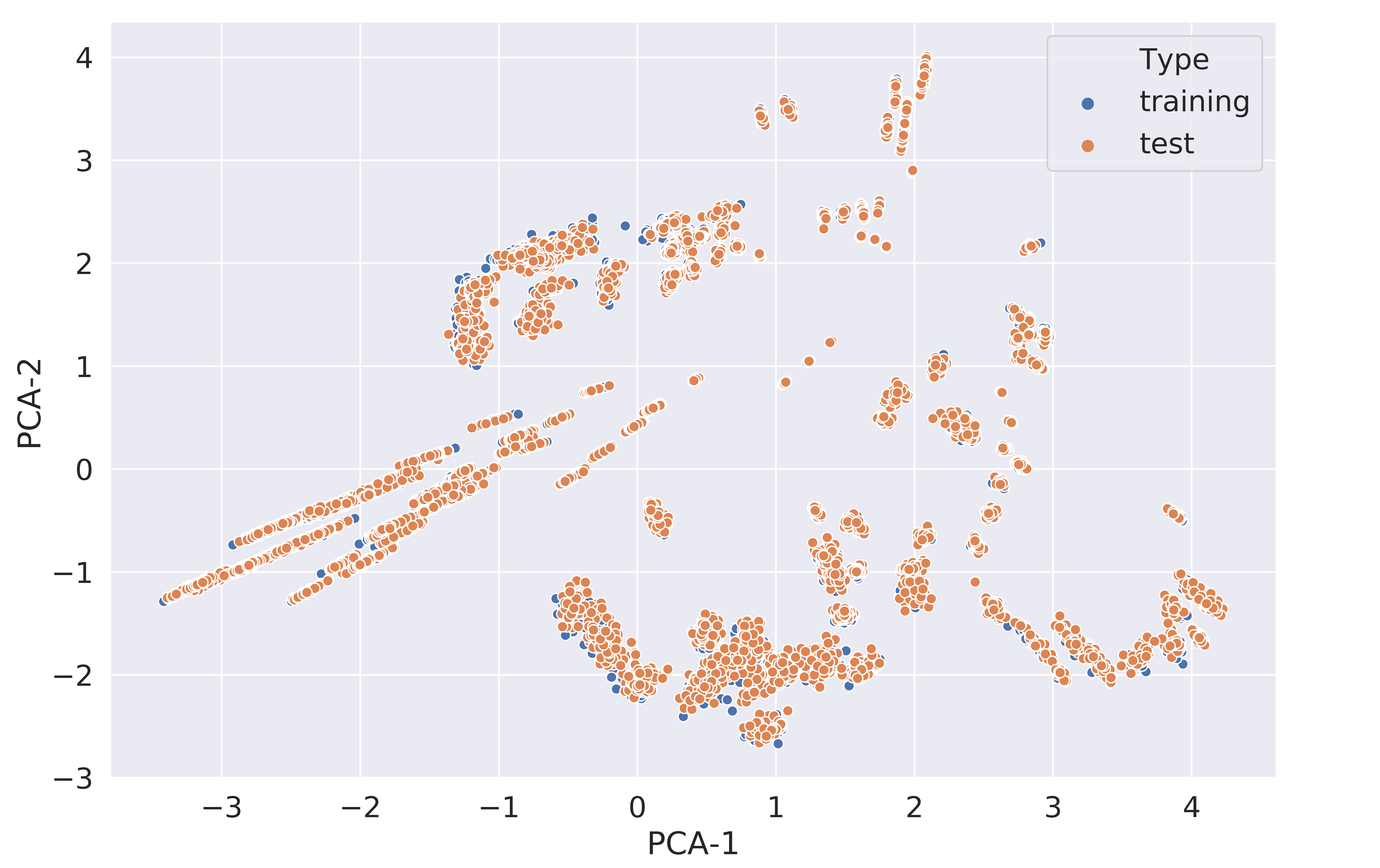}
    \caption{Dimensionality reduction with principal component analysis (PCA) \\
            of Gaussian features of training and test sets.}
    \label{fig:pcacloseness}
\end{figure}

It is worth noting that data points in the ANI-1 dataset are obtained through
normal model sampling (NMS) around the equilibrium geometry. Therefore,
Gaussian features of atoms are very close to each other in the feature space.
To prove this hypothesis we carried out a principal component analysis (PCA)
dimensionality reduction of the Gaussian features of the training and test
sets. In Figure~\ref{fig:pcacloseness}, we can see how the training and test
data points are close in the lower PCA dimensional space making it easy for
the model to correctly predict them.

\subsection{Kernel Ridge Regression with Coulomb Matrix Features}

In this demonstration, we train a kernel ridge regression (KRR) model for the
task of predicting atomization energies with the QM7 data set. Instead of the
\lstinline{Potentials} class, we will use a modular approach to show the
flexibility of \textsf{ML4Chem}. We proceed, as shown in
Section~\ref{sec:datahandling}, to the instantiation of the
\lstinline{Data()} class with the \lstinline{qmt7.traj} trajectory file for
the purpose of \lstinline{"training"}. The training set and targets are
obtained with the \lstinline{.get_data()} class method to interoperate with
\textsf{ML4Chem} (see Listing~\ref{lst:krrtraining}).

We use the rows of the Coulomb matrix features~\cite{Collins2018} as
atomistic features vectors. The training set is featurized by instantiating
the \lstinline{CoulombMatrix} class implemented using the DScribe package and
calling the \lstinline{.calculate()} method. In the next step, the
\lstinline{KernelRidge} class is instantiated with keyword arguments to set
training data, batch size, kernel function (radial basis function, or RBF),
and a sigma value. The model is prepared with features and reference space.
In the last step, we call the \lstinline{.train()} method to fit the KRR
algorithm passing as arguments the training set, targets, and
\lstinline{Data} objects and the model's parameters are saved to disk.

\begin{figure}[ht]
    \centering
    \includegraphics[width=11.0cm]{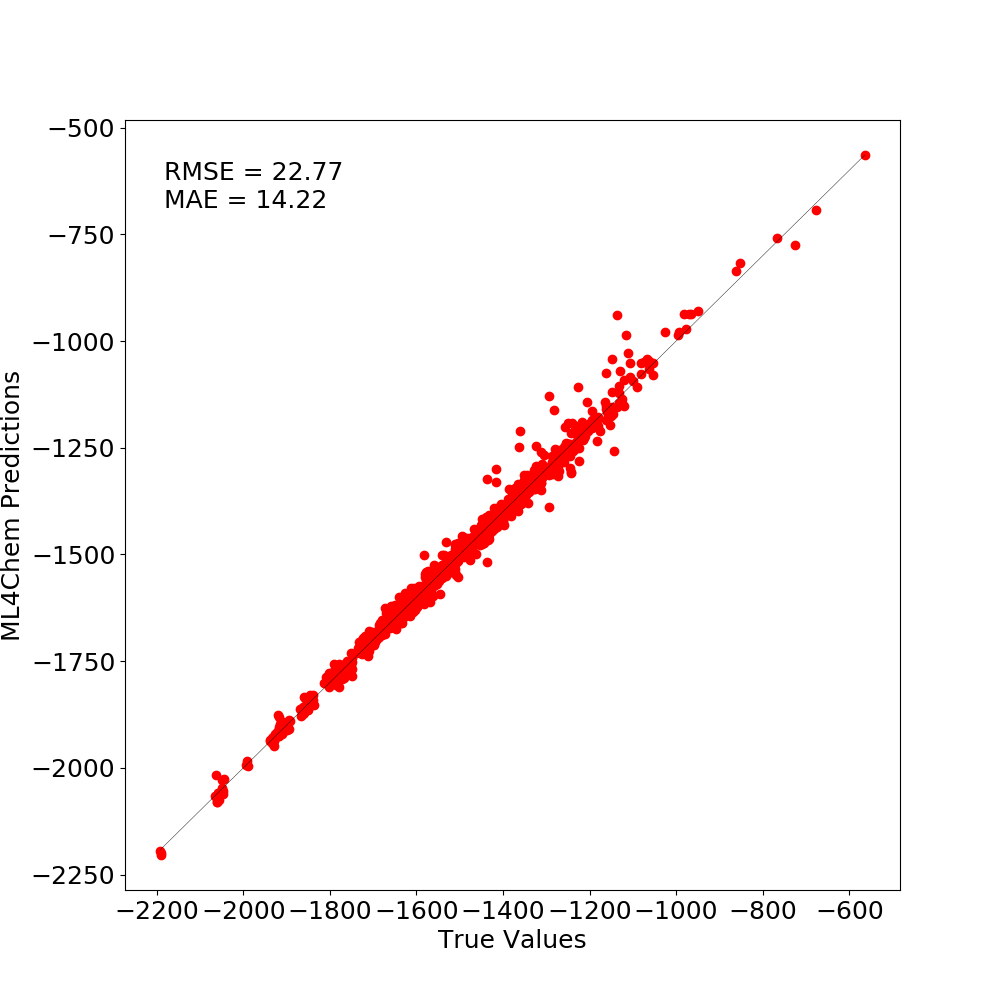}
    \caption{Parity plot of kernel ridge regression predictions over $1000$
    molecules \\ in the QM7 data set. Units are in kcal/mol.}
    \label{fig:krrcoulombqm7}
\end{figure}

Now, we can proceed to load this model and predict unknown data points (see
Listing~\ref{lst:krrinference}). The results are shown in a parity plot in
Figure~\ref{fig:krrcoulombqm7} with the root-mean-squared (RMSE) and mean
squared (MAE) error metrics. This model showed an MAE of $14.2$ kcal/mol,
compared to $9.9$ kcal/mol when using the same RBF kernel on the Coulomb
matrix sorted eigenspectrum in Ref.~\cite{Rupp2012}. This significant
difference is expected because, for this demonstration, we determined sigma
to be the average of the euclidean distance of the Coulomb matrix atomic
feature vectors. This is not an exhaustive way of determining this
hyperparameter, and in practice, one has to rely on k-fold cross-validation
to find the best value that fits the data of interest. Also, the number of
training set data points in our demonstration is smaller than the used in
Ref~\cite{Rupp2012}.

\newpage
\begin{lstlisting}[caption={Training a kernel ridge
regression algorithm in \textsf{ML4Chem}.},
label={lst:krrtraining},language=Python]
from ml4chem.data.handler import Data
from ase.io import read
from ml4chem.utils import logger
from ml4chem.atomistic.features import CoulombMatrix
from ml4chem.atomistic.models import KernelRidge
from ml4chem.atomistic import Atomistic
from dask.distributed import Client, LocalCluster


def run():
    # Use 500 molecules for a total of 7727 feature vectors
    n_molecules, batch_size = 500, 10

    # Start a logger object to write to file
    logger("experiments.log")

    # Read training data
    training = read("training_images.traj", index="0:{}:1".format(n_molecules))

    # Prepare Data object
    data = Data(training, purpose="training")
    training, targets = data.get_data(purpose="training")

    # Featurization using Coulomb Matrix 
    n_atoms_max = max(data.atoms_per_image)
    cm = CoulombMatrix(n_atoms_max=n_atoms_max, batch_size=batch_size)
    features, reference_space = cm.calculate(training, data=data, svm=True, purpose="training")

    # Instatiate model, prepare and train. sigma is set to average Euclidean
    # distance of feature vectors
    krr = KernelRidge(
        trainingimages="training_images.traj", batch_size=batch_size, kernel="rbf", 
        sigma=26.808046418478668
    )

    krr.prepare_model(features, reference_space, data=data, purpose="training")
    krr.train(training, targets, data=data)

    # Save model to disk
    Atomistic.save(krr, features=cm, path="krr/", label="publication")


if __name__ == "__main__":
    cluster = LocalCluster(n_workers=15, threads_per_worker=1)
    client = Client(cluster)
    run()

\end{lstlisting}

\newpage
\begin{lstlisting}[caption={Inference using trained
kernel ridge regression \textsf{ML4Chem} parameters.}, label={lst:krrinference},language=Python]
from ase.io import read
from ml4chem.utils import logger
from ml4chem.atomistic import Atomistic
from dask.distributed import Client, LocalCluster
from ml4chem.data.visualization import parity
import pandas as pd


def run():
    # Use 1000 molecules from the test set
    n_molecules = 1000

    # Start a logger object to write to file
    logger("inference.log")

    # Read test data
    calc = Atomistic.load(model="krr/publication.ml4c", params="krr/publication.params")

    # Set the reference space
    calc.reference_space = "features.db"

    # Compute energies
    energies = []
    trues = []

    for index, atoms in enumerate(test):
        energy = calc.get_potential_energy(atoms)
        true = atoms.get_potential_energy()
        print(true, energy)
        trues.append(true)
        energies.append(energy)

    df = pd.DataFrame(
        {"ML4Chem Energies": energies, "True Values": trues}
    )
    df.to_pickle("inference_results.pkl")

    parity(
        energies, trues, scores=True, filename="parity_inference.png"
    )


if __name__ == "__main__":
    cluster = LocalCluster(n_workers=15,threads_per_worker=1)
    client = Client(cluster)
    run()
\end{lstlisting}

\section{Conclusions}

We presented the \lstinline{atomistic} module of \textsf{ML4Chem}, an open-source
software package aiming to ease the deployment and implementation of ML
models in chemistry and materials science. Its structure is designed with
strict modularity and defined in such a way that each of its parts can be
used as standalone programs. \textsf{ML4Chem} provides all needed methods and
tools for an ML discovery pipeline -- that is to go from raw data to
inference and visualization. We showed with code snippets and demonstration
cases what can be achieved with the core \lstinline{atomistic} modules, and
the intended intuitiveness derived from the use of UX design rules. These
attributes make \textsf{ML4Chem} a potential platform for the implementation
of new models, targeted to both non-experts and expert users.

For future development directions, we plan to extend the support of other
input data formats of the~ \lstinline{atomistic} module beyond the atomic
simulation environment (ASE); introduce a \lstinline{geometric} module in
\textsf{ML4Chem} to work with geometric deep learning; addition of
convolutional neural networks for both the \lstinline{atomistic} and
\lstinline{geometric} modules; transfer learning; and implementation of
active learning protocols.

\section{Acknowledgement}

We acknowledge the Laboratory Directed Research and Development of Lawrence
Berkeley National Laboratory for funding under the project ID: 105702. MEK
acknowledges Thom Popovici (LBL) for fruitful discussions related to
high-performance computing parallelism, Scott Sievert (UW–Madison) and
Matthew Rocklin (NVIDIA) for helpful advice on the Dask library.

\bibliographystyle{unsrt} 
\bibliography{main.bib}

\begin{thebibliography}{10}

\bibitem{Goodfellow2016}
Ian Goodfellow, Yoshua Bengio, and Aaron Courville.
\newblock {\em Deep Learning}.
\newblock MIT Press, 2016.

\bibitem{TensorFlow}
Martin Abadi, Paul Barham, Jianmin Chen, Zhifeng Chen, Andy Davis, Jeffrey
  Dean, Matthieu Devin, Sanjay Ghemawat, Geoffrey Irving, Michael Isard,
  Manjunath Kudlur, Josh Levenberg, Rajat Monga, Sherry Moore, Derek~G. Murray,
  Benoit Steiner, Paul Tucker, Vijay Vasudevan, Pete Warden, Martin Wicke, Yuan
  Yu, and Xiaoqiang Zheng.
\newblock Tensorflow: A system for large-scale machine learning.
\newblock In {\em 12th USENIX Symposium on Operating Systems Design and
  Implementation (OSDI 16)}, pages 265--283, 2016.

\bibitem{NEURIPS2019_9015}
Adam Paszke, Sam Gross, Francisco Massa, Adam Lerer, James Bradbury, Gregory
  Chanan, Trevor Killeen, Zeming Lin, Natalia Gimelshein, Luca Antiga, Alban
  Desmaison, Andreas Kopf, Edward Yang, Zachary DeVito, Martin Raison, Alykhan
  Tejani, Sasank Chilamkurthy, Benoit Steiner, Lu~Fang, Junjie Bai, and Soumith
  Chintala.
\newblock {PyTorch: An Imperative Style, High-Performance Deep Learning
  Library}.
\newblock In H~Wallach, H~Larochelle, A~Beygelzimer,
  F~d$\backslash$textquotesingle Alch{\'{e}}-Buc, E~Fox, and R~Garnett,
  editors, {\em Advances in Neural Information Processing Systems 32}, pages
  8024--8035. Curran Associates, Inc., 2019.

\bibitem{Artrith2012}
Nongnuch Artrith and J{\"{o}}rg Behler.
\newblock {High-dimensional neural network potentials for metal surfaces: A
  prototype study for copper}.
\newblock {\em Physical Review B}, 85(4):045439, jan 2012.

\bibitem{Artrith2011}
Nongnuch Artrith, Tobias Morawietz, and J{\"{o}}rg Behler.
\newblock {High-dimensional neural-network potentials for multicomponent
  systems: Applications to zinc oxide}.
\newblock {\em Physical Review B}, 83(15):153101, apr 2011.

\bibitem{Behler2010}
J{\"{o}}rg Behler.
\newblock {Neural network potential-energy surfaces for atomistic simulations}.
\newblock In {\em Chemical Modelling}, pages 1--41. Royal Society of Chemistry,
  Cambridge, 2010.

\bibitem{Behler2007}
J{\"{o}}rg Behler and Michele Parrinello.
\newblock {Generalized Neural-Network Representation of High-Dimensional
  Potential-Energy Surfaces}.
\newblock {\em Physical Review Letters}, 98(14):146401, apr 2007.

\bibitem{Pilania2013}
Ghanshyam Pilania, Chenchen Wang, Xun Jiang, Sanguthevar Rajasekaran, and
  Ramamurthy Ramprasad.
\newblock {Accelerating materials property predictions using machine learning}.
\newblock {\em Scientific Reports}, 3(1):2810, dec 2013.

\bibitem{Toth2019}
Peter Toth, Danilo~Jimenez Rezende, Andrew Jaegle, S{\'{e}}bastien
  Racani{\`{e}}re, Aleksandar Botev, and Irina Higgins.
\newblock {Hamiltonian Generative Networks}.
\newblock pages 1--17, sep 2019.

\bibitem{Schutt2019a}
K.~T. Sch{\"{u}}tt, M.~Gastegger, A.~Tkatchenko, K.-R. M{\"{u}}ller, and R.~J.
  Maurer.
\newblock {Unifying machine learning and quantum chemistry with a deep neural
  network for molecular wavefunctions}.
\newblock {\em Nature Communications}, 10(1):5024, dec 2019.

\bibitem{Gomez-Bombarelli2018}
Rafael G{\'{o}}mez-Bombarelli, Jennifer~N. Wei, David Duvenaud,
  Jos{\'{e}}~Miguel Hern{\'{a}}ndez-Lobato, Benjam{\'{i}}n
  S{\'{a}}nchez-Lengeling, Dennis Sheberla, Jorge Aguilera-Iparraguirre,
  Timothy~D. Hirzel, Ryan~P. Adams, and Al{\'{a}}n Aspuru-Guzik.
\newblock {Automatic Chemical Design Using a Data-Driven Continuous
  Representation of Molecules}.
\newblock {\em ACS Central Science}, 4(2):268--276, feb 2018.

\bibitem{Sanchez-Lengeling2019}
Benjamin Sanchez-Lengeling, Jennifer~N. Wei, Brian~K. Lee, Richard~C. Gerkin,
  Al{\'{a}}n Aspuru-Guzik, and Alexander~B. Wiltschko.
\newblock {Machine Learning for Scent: Learning Generalizable Perceptual
  Representations of Small Molecules}.
\newblock oct 2019.

\bibitem{Hutson2018}
Matthew Hutson.
\newblock {Artificial intelligence faces reproducibility crisis}.
\newblock {\em Science}, 359(6377):725 LP -- 726, feb 2018.

\bibitem{Stupple2019}
Aaron Stupple, David Singerman, and Leo~Anthony Celi.
\newblock {The reproducibility crisis in the age of digital medicine}.
\newblock {\em npj Digital Medicine}, 2(1):2, dec 2019.

\bibitem{deepchem}
Democratizing deep-learning for drug discovery, quantum chemistry, materials
  science and biology.
\newblock \url{https://github.com/deepchem/deepchem}, 2016.

\bibitem{chemml2019}
Mojtaba Haghighatlari, Gaurav Vishwakarma, Doaa Altarawy, Ramachandran
  Subramanian, Bhargava~U. Kota, Aditya Sonpal, Srirangaraj Setlur, and
  Johannes Hachmann.
\newblock Chemml: A machine learning and informatics program package for the
  analysis, mining, and modeling of chemical and materials data.
\newblock {\em WIREs Computational Molecular Science}, n/a(n/a):e1458.

\bibitem{Li2015}
Zhenwei Li, James~R. Kermode, and Alessandro {De Vita}.
\newblock {Molecular Dynamics with On-the-Fly Machine Learning of
  Quantum-Mechanical Forces}.
\newblock {\em Physical Review Letters}, 114(9):096405, mar 2015.

\bibitem{Botu2015b}
Venkatesh Botu and Rampi Ramprasad.
\newblock {Adaptive machine learning framework to accelerate ab initio
  molecular dynamics}.
\newblock {\em International Journal of Quantum Chemistry}, 115(16):1074--1083,
  aug 2015.

\bibitem{Chmiela2017}
Stefan Chmiela, Alexandre Tkatchenko, Huziel~E. Sauceda, Igor Poltavsky,
  Kristof~T. Sch{\"{u}}tt, and Klaus-Robert M{\"{u}}ller.
\newblock Machine learning of accurate energy-conserving molecular force
  fields.
\newblock {\em Science Advances}, 3(5):e1603015, may 2017.

\bibitem{Li2018a}
Linglong Li, Yaodong Yang, Dawei Zhang, Zuo~Guang Ye, Stephen Jesse, Sergei~V.
  Kalinin, and Rama~K. Vasudevan.
\newblock {Machine learning-enabled identification of material phase
  transitions based on experimental data: Exploring collective dynamics in
  ferroelectric relaxors}.
\newblock {\em Science Advances}, 4(3), 2018.

\bibitem{Botu2015}
Venkatesh Botu and R.~Ramprasad.
\newblock {Learning scheme to predict atomic forces and accelerate materials
  simulations}.
\newblock {\em Physical Review B}, 92(9):094306, sep 2015.

\bibitem{Botu2017a}
Venkatesh Botu, R.~Batra, J.~Chapman, and R.~Ramprasad.
\newblock {Machine Learning Force Fields: Construction, Validation, and
  Outlook}.
\newblock {\em The Journal of Physical Chemistry C}, 121(1):511--522, 2017.

\bibitem{Peterson2016a}
Andrew~A. Peterson.
\newblock Acceleration of saddle-point searches with machine learning.
\newblock {\em The Journal of Chemical Physics}, 145(7):074106, aug 2016.

\bibitem{Ghasemi2015}
S.~Alireza Ghasemi, Albert Hofstetter, Santanu Saha, and Stefan Goedecker.
\newblock Interatomic potentials for ionic systems with density functional
  accuracy based on charge densities obtained by a neural network.
\newblock {\em Physical Review B}, 92(4):045131, jul 2015.

\bibitem{Faraji2017}
Somayeh Faraji, S.~Alireza Ghasemi, Samare Rostami, Robabe Rasoulkhani, Bastian
  Schaefer, Stefan Goedecker, and Maximilian Amsler.
\newblock {High accuracy and transferability of a neural network potential
  through charge equilibration for calcium fluoride}.
\newblock {\em Physical Review B}, 95(10):104105, mar 2017.

\bibitem{VanderWalt2011}
St{\'{e}}fan {Van Der Walt}, S~Chris Colbert, and Ga{\"{e}}l Varoquaux.
\newblock {The NumPy array: A structure for efficient numerical computation}.
\newblock {\em Computing in Science and Engineering}, 13(2):22--30, 2011.

\bibitem{Millman2011}
K.~Jarrod Millman and Michael Aivazis.
\newblock {Python for scientists and engineers}.
\newblock {\em Computing in Science and Engineering}, 13(2):9--12, 2011.

\bibitem{2019arXiv190710121V}
Pauli Virtanen, Ralf Gommers, Travis~E Oliphant, Matt Haberland, Tyler Reddy,
  David Cournapeau, Evgeni Burovski, Pearu Peterson, Warren Weckesser, Jonathan
  Bright, St{\'{e}}fan~J van~der Walt, Matthew Brett, Joshua Wilson, K~{Jarrod
  Millman}, Nikolay Mayorov, Andrew~R.{\~{}}J. Nelson, Eric Jones, Robert Kern,
  Eric Larson, C~J Carey, $\backslash$.Ilhan Polat, Yu~Feng, Eric~W Moore, Jake
  Vand~erPlas, Denis Laxalde, Josef Perktold, Robert Cimrman, Ian Henriksen,
  E.{\~{}}A. Quintero, Charles~R Harris, Anne~M Archibald, Ant{\^{o}}nio~H
  Ribeiro, Fabian Pedregosa, Paul van Mulbregt, and SciPy 1.~0 Contributors.
\newblock {SciPy 1.0--Fundamental Algorithms for Scientific Computing in
  Python}.
\newblock {\em arXiv e-prints}, page arXiv:1907.10121, jul 2019.

\bibitem{DaskDevelopmentTeam2016}
{Dask Development Team}.
\newblock {Dask: Library for dynamic task scheduling}, 2016.

\bibitem{HjorthLarsen2017}
Ask {Hjorth Larsen}, Jens {J{\o}rgen Mortensen}, Jakob Blomqvist, Ivano~E
  Castelli, Rune Christensen, Marcin Du{\l}ak, Jesper Friis, Michael~N Groves,
  Bj{\o}rk Hammer, Cory Hargus, Eric~D Hermes, Paul~C Jennings, Peter {Bjerre
  Jensen}, James Kermode, John~R Kitchin, Esben {Leonhard Kolsbjerg}, Joseph
  Kubal, Kristen Kaasbjerg, Steen Lysgaard, J{\'{o}}n {Bergmann Maronsson},
  Tristan Maxson, Thomas Olsen, Lars Pastewka, Andrew Peterson, Carsten
  Rostgaard, Jakob Schi{\o}tz, Ole Sch{\"{u}}tt, Mikkel Strange, Kristian~S
  Thygesen, Tejs Vegge, Lasse Vilhelmsen, Michael Walter, Zhenhua Zeng, and
  Karsten~W Jacobsen.
\newblock The atomic simulation environment—a python library for working with
  atoms.
\newblock {\em Journal of Physics: Condensed Matter}, 29(27):273002, jul 2017.

\bibitem{Hanwell2017}
Marcus~D. Hanwell, Wibe~A. {De Jong}, and Christopher~J. Harris.
\newblock {Open chemistry: RESTful web APIs, JSON, NWChem and the modern web
  application}.
\newblock {\em Journal of Cheminformatics}, 9(1):1--10, 2017.

\bibitem{Smith2017}
J.~S. Smith, O.~Isayev, and A.~E. Roitberg.
\newblock Ani-1: an extensible neural network potential with dft accuracy at
  force field computational cost.
\newblock {\em Chem. Sci.}, 8(4):3192--3203, 2017.

\bibitem{Ong2013}
Shyue~Ping Ong, William~Davidson Richards, Anubhav Jain, Geoffroy Hautier,
  Michael Kocher, Shreyas Cholia, Dan Gunter, Vincent~L. Chevrier, Kristin~A.
  Persson, and Gerbrand Ceder.
\newblock {Python Materials Genomics (pymatgen): A robust, open-source python
  library for materials analysis}.
\newblock {\em Computational Materials Science}, 68:314--319, feb 2013.

\bibitem{Rupp2012}
Matthias Rupp, Alexandre Tkatchenko, Klaus-Robert M{\"{u}}ller, and O.~Anatole
  von Lilienfeld.
\newblock {Fast and Accurate Modeling of Molecular Atomization Energies with
  Machine Learning}.
\newblock {\em Physical Review Letters}, 108(5):058301, jan 2012.

\bibitem{Bartok2013}
Albert~P. Bart{\'{o}}k, Risi Kondor, and G{\'{a}}bor Cs{\'{a}}nyi.
\newblock {On representing chemical environments}.
\newblock {\em Physical Review B}, 87(18):184115, may 2013.

\bibitem{Huo2017a}
Haoyan Huo and Matthias Rupp.
\newblock {Unified Representation of Molecules and Crystals for Machine
  Learning}.
\newblock apr 2017.

\bibitem{Himanen2019}
Lauri Himanen, Marc O.~J. J{\"{a}}ger, Eiaki~V. Morooka, Filippo~Federici
  Canova, Yashasvi~S. Ranawat, David~Z. Gao, Patrick Rinke, and Adam~S. Foster.
\newblock {DScribe: Library of Descriptors for Machine Learning in Materials
  Science}.
\newblock apr 2019.

\bibitem{Khorshidi2016}
Alireza Khorshidi and Andrew~A. Peterson.
\newblock {Amp : A modular approach to machine learning in atomistic
  simulations}.
\newblock {\em Computer Physics Communications}, 207:310--324, oct 2016.

\bibitem{Behler2015}
J{\"{o}}rg Behler.
\newblock {Constructing high-dimensional neural network potentials: A tutorial
  review}.
\newblock {\em International Journal of Quantum Chemistry}, 115(16):1032--1050,
  aug 2015.

\bibitem{scikit-learn}
F~Pedregosa, G~Varoquaux, A~Gramfort, V~Michel, B~Thirion, O~Grisel, M~Blondel,
  P~Prettenhofer, R~Weiss, V~Dubourg, J~Vanderplas, A~Passos, D~Cournapeau,
  M~Brucher, M~Perrot, and E~Duchesnay.
\newblock {Scikit-learn: Machine Learning in {\{}P{\}}ython}.
\newblock {\em Journal of Machine Learning Research}, 12:2825--2830, 2011.

\bibitem{Borsboom2003}
Denny Borsboom, Gideon~J. Mellenbergh, and Jaap {Van Heerden}.
\newblock {The Theoretical Status of Latent Variables}.
\newblock {\em Psychological Review}, 110(2):203--219, 2003.

\bibitem{Fu2019}
Hao Fu, Chunyuan Li, Xiaodong Liu, Jianfeng Gao, Asli Celikyilmaz, and Lawrence
  Carin.
\newblock {Cyclical Annealing Schedule: A Simple Approach to Mitigating}.
\newblock pages 240--250, 2019.

\bibitem{ricci2011introduction}
Francesco Ricci, Lior Rokach, and Bracha Shapira.
\newblock {Introduction to recommender systems handbook}.
\newblock In {\em Recommender systems handbook}, pages 1--35. Springer, 2011.

\bibitem{Tang2018}
Yu-Hang Tang and Wibe~A. de~Jong.
\newblock {Prediction of Atomization Energy Using Graph Kernel and Active
  Learning}.
\newblock pages 1--19, oct 2018.

\bibitem{geisser2017predictive}
Seymour Geisser.
\newblock {\em Predictive inference}.
\newblock Routledge, 2017.

\bibitem{Burkov2019}
Andriy Burkov.
\newblock {\em The Hundred-Page Machine Learning Book}.
\newblock Kindle Direct Publishing, 1 edition, 2019.

\bibitem{P.Murphy2012}
Kevin {P. Murphy}.
\newblock {\em Machine Learning}.
\newblock Springer-Verlag, Berlin/Heidelberg, 2012.

\bibitem{Marsland2015}
Stephen Marsland.
\newblock {\em Machine learning: an algorithmic perspective}.
\newblock Chapman and Hall/CRC, dec 2014.

\bibitem{Sowmya2004}
A.~Sowmya.
\newblock {The anisotropic Gaussian kernel for SVM classification of HRCT
  images of the lung}.
\newblock {\em Proceedings of the 2004 Intelligent Sensors, Sensor Networks and
  Information Processing Conference, 2004.}, pages 439--444, 2004.

\bibitem{Rupp2015}
Matthias Rupp.
\newblock {Machine learning for quantum mechanics in a nutshell}.
\newblock {\em International Journal of Quantum Chemistry}, 115(16):1058--1073,
  aug 2015.

\bibitem{Bartok2010}
Albert~P. Bart{\'{o}}k, Mike~C. Payne, Risi Kondor, and G{\'{a}}bor
  Cs{\'{a}}nyi.
\newblock {Gaussian Approximation Potentials: The Accuracy of Quantum
  Mechanics, without the Electrons}.
\newblock {\em Physical Review Letters}, 104(13):136403, apr 2010.

\bibitem{Bartok2015}
Albert~P. Bart{\'{o}}k and G{\'{a}}bor Cs{\'{a}}nyi.
\newblock {Gaussian approximation potentials: A brief tutorial introduction}.
\newblock {\em International Journal of Quantum Chemistry}, 115(16):1051--1057,
  aug 2015.

\bibitem{Lemke2019}
Tobias Lemke and Christine Peter.
\newblock {EncoderMap: Dimensionality Reduction and Generation of Molecule
  Conformations}.
\newblock {\em Journal of Chemical Theory and Computation}, 15(2):1209--1215,
  feb 2019.

\bibitem{Makhzani2013}
Alireza Makhzani and Brendan Frey.
\newblock {k-Sparse Autoencoders}.
\newblock dec 2013.

\bibitem{Arpit2015}
Devansh Arpit, Yingbo Zhou, Hung Ngo, and Venu Govindaraju.
\newblock {Why Regularized Auto-Encoders learn Sparse Representation?}
\newblock may 2015.

\bibitem{Kingma2013}
Diederik~P Kingma and Max Welling.
\newblock {Auto-Encoding Variational Bayes}.
\newblock {\em arXiv preprint arXiv:1312.6114}, dec 2013.

\bibitem{7926714}
X~Hou, L~Shen, K~Sun, and G~Qiu.
\newblock {Deep Feature Consistent Variational Autoencoder}.
\newblock In {\em 2017 IEEE Winter Conference on Applications of Computer
  Vision (WACV)}, pages 1133--1141, mar 2017.

\bibitem{Zhao2017}
Shengjia Zhao, Jiaming Song, and Stefano Ermon.
\newblock {Towards Deeper Understanding of Variational Autoencoding Models}.
\newblock feb 2017.

\bibitem{kullback1951information}
Solomon Kullback and Richard~A Leibler.
\newblock On information and sufficiency.
\newblock {\em The annals of mathematical statistics}, 22(1):79--86, 1951.

\bibitem{pascanu2012understanding}
Razvan Pascanu, Tomas Mikolov, and Yoshua Bengio.
\newblock {Understanding the exploding gradient problem}.
\newblock {\em CoRR, abs/1211.5063}, 2, 2012.

\bibitem{glorot2011deep}
Xavier Glorot, Antoine Bordes, and Yoshua Bengio.
\newblock {Deep sparse rectifier neural networks}.
\newblock In {\em Proceedings of the fourteenth international conference on
  artificial intelligence and statistics}, pages 315--323, 2011.

\bibitem{Kingma2014}
Diederik~P. Kingma and Jimmy~Lei Ba.
\newblock {Adam: A method for stochastic optimization}.
\newblock In {\em 3rd International Conference on Learning Representations,
  ICLR 2015 - Conference Track Proceedings}, dec 2015.

\bibitem{robbins1951stochastic}
Herbert Robbins and Sutton Monro.
\newblock {A stochastic approximation method}.
\newblock {\em The annals of mathematical statistics}, pages 400--407, 1951.

\bibitem{duchi2011adaptive}
John Duchi, Elad Hazan, and Yoram Singer.
\newblock {Adaptive subgradient methods for online learning and stochastic
  optimization}.
\newblock {\em Journal of Machine Learning Research}, 12(Jul):2121--2159, 2011.

\bibitem{Hunter:2007}
J~D Hunter.
\newblock {Matplotlib: A 2D graphics environment}.
\newblock {\em Computing in Science {\&} Engineering}, 9(3):90--95, 2007.

\bibitem{plotly}
Plotly~Technologies Inc.
\newblock Collaborative data science.
\newblock 2015.

\bibitem{rupp2012data}
M.~Rupp, A.~Tkatchenko, K.-R. M\"uller, and O.~A. von Lilienfeld.
\newblock Fast and accurate modeling of molecular atomization energies with
  machine learning.
\newblock {\em Physical Review Letters}, 108:058301, 2012.

\bibitem{blum2009}
L.~C. Blum and J.-L. Reymond.
\newblock 970 million druglike small molecules for virtual screening in the
  chemical universe database {GDB-13}.
\newblock {\em J. Am. Chem. Soc.}, 131:8732, 2009.

\bibitem{Collins2018}
Christopher~R. Collins, Geoffrey~J. Gordon, O.~Anatole {Von Lilienfeld}, and
  David~J. Yaron.
\newblock {Constant size descriptors for accurate machine learning models of
  molecular properties}.
\newblock {\em J. Chem. Phys.}, 148(24), jun 2018.

\end{thebibliography}

\end{document}